\documentclass[11pt,a4paper]{article}
\pdfoutput=1
\usepackage{jheppub}

\newcommand{\rar}{\rightarrow}

\newcommand{\gsim}{ \lower .75ex \hbox{$\sim$} \llap{\raise .27ex \hbox{$>$}} }
\newcommand{\lsim}{ \lower .75ex \hbox{$\sim$} \llap{\raise .27ex \hbox{$<$}} }

\long\def\nicefootnote[#1]#2{\begingroup%
\def\thefootnote{\arabic{footnote}}\footnote{#2}\endgroup}

\def\be{\begin{equation}}
\def\ee{\end{equation}}
\def\bea{\begin{eqnarray}}
\def\eea{\end{eqnarray}}

\begin{document}

\title{Bose-Fermi competition in holographic metals\\[.3in]}
\author{Yan Liu,}
\author{Koenraad Schalm,}
\author{Ya-Wen Sun,} 
\author{Jan Zaanen}

\affiliation{
~\\
\mbox{Institute Lorentz for Theoretical Physics, Leiden University}\\
\mbox{P.O. Box 9506, Leiden 2300RA, The Netherlands}}

\emailAdd{liu, kschalm, sun, jan@lorentz.leidenuniv.nl.}



\abstract{We study the holographic dual of a finite density system with both bosonic and fermionic degrees of freedom. There is no evidence for a universal bose-dominated ground state. Instead, depending on the relative conformal weights the preferred groundstate is either pure AdS-Reissner-Nordstrom, a holographic superconductor, an electron star, or a novel mixed state that is best characterized as a hairy electron star. }


\maketitle

\setcounter{tocdepth}{2}
\section{Introduction}

It is a rule of thumb that at any given finite density bosons always win from fermions. Many bosons can coherently occupy the groundstate whereas the Gibbs potential gain decreases with each additional fermion due to the Pauli principle. At the same time it is not difficult to construct a system where this notion does not hold. A relativistic system with massive bosons but massless fermions, will first occupy the fermionic modes until the chemical potential reaches the mass of boson, or if there is an incredibly large number of distinguishable degenerate fermions, Pauli blocking is not relevant.
 
Strongly coupled system with no clear particle spectrum, {\it e.g.} conformal field theories, are another system where the validity of this rule is not obvious. Using the insight offered by the AdS/CFT correspondence, we study here combined Bose-Fermi systems at finite density.\footnote{Holographic bosonic competing systems have been studied in {\it e.g.} \cite{Basu:2010fa, {Donos:2012yu}, {Musso:2013ija}, {Cai:2013wma}}.} For each separately AdS/CFT has already given us some remarkable if not revolutionary insights: one can describe the condensation of strongly coupled bosonic systems at finite density with order parameter dimensions that are far beyond perturbation theory \cite{Hartnoll:2008vx, {Hartnoll:2008kx},{Gubser:2009cg},{Horowitz:2009ij}}. Fermionic AdS/CFT systems naturally describe non-Fermi-liquid states \cite{Liu:2009dm,{Cubrovic:2009ye},{Faulkner:2009wj}}. The origin of this exotic physics can be traced to the interplay between the charged sector exposed by the chemical potential and a large neutral critical sector that survives in the deep IR \cite{Faulkner:2010tq}. Standard generic condensed matter wisdoms are recovered when this sector is lifted. This removes the strongly coupled particle-less physics from the deep IR. 

We shall stay within the confines of the standard AdS/CFT set-up and
study this Bose-Fermi competition in the strongly coupled regime. For
the fermions we shall take a conventional fluid approximation in the
bulk (Sec. \ref{sectionsetup}). This is known to correspond to a large number of distinct Fermi
surfaces in the dual theory \cite{Hartnoll:2010gu, {Hartnoll:2011dm},{Iqbal:2011in},{Cubrovic:2011xm}}. It already indicates that the simple
Bose-Fermi competition rule might not hold. Indeed we find that,
depending on the charges and conformal dimensions of the bosonic and
fermionic operators, a mixed regime exists (Sec. \ref{zerotsol}). The gravitational dual to this regime is an electron star with charged scalar hair and we call this a hairy electron star solution. 
In the 
section \ref{phasediag} we explore the phase diagram of this
system at zero temperature as a function of the scaling dimensions of
the bose and fermi fields respectively. We find that each of the four phases
dominates a distinct region in the phase diagram.
Our conclusions and discussion of the result are in section \ref{condis}.
\vspace{.5cm}\\
{\bf Note added:} As we were finalizing our paper, we were informed that F. Nitti, G. Policastro and T. Vanel have obtained similar results 
\cite{Nitti:2013xaa}. 

\section{Set-up} 
\label{sectionsetup}

The gravity Lagrangian encoding the strongly coupled field theory we
consider is 3+1 dimensional AdS-Einstein-Maxwell theory with a charged
massive scalar and a charged massive fermion. We only consider
renormalizable interactions in the bulk, corresponding to the most
relevant operators in the large $N$ expansion of the field theory. For
generic charges, when $q_b\neq 2q_f$ and $q_b\neq 0$, there is no such
renormalizable Yukawa coupling and there is no direct interaction
between the bosons and fermions. In the context that we are interested
in, the most general (parity conserving) gravity Lagrangian is therefore
\bea \mathcal{L}&=&\frac{1}{2\kappa^2}\bigg(R+\frac{6}{L^2}\bigg)-\frac{1}{4e^2}F_{\mu\nu}F^{\mu\nu}
-|(\partial_\mu-iq_bA_\mu)\phi|^2-V(|\phi|)-i\bar\Psi(\Gamma^\mu  \mathcal{D}_\mu-m_f)\Psi, \nonumber\\
\eea
where
\be V(\phi)=\frac{u}{2L^2}\big(|\phi|^2+\frac{m_b^2L^2}{u}\big)^2-\frac{m_b^4L^2}{2u},\ee
\be\bar{\Psi}=\Psi^{\dagger}\Gamma^{\underline{t}}, ~~~\mathcal{D}_\mu=\partial_\mu+\frac{1}{4}\omega_{ab \mu}\Gamma^{ab}-iq_fA_\mu.
\ee
Here  $m_b$, $q_b$, and $m_f$, $q_f$, denote the mass and charge of
the bosons and the fermions respectively, and $\kappa$, $e$, $u$ are
the Newton constant, the Maxwell coupling constant and the $\phi ^4$
coupling constant. Rescaling $A_{\mu} \rar e A_{\mu}$ shows that the
action only depends on the combinations $eq_b$ and $eq_f$ and we will
use this to fix $q_f=1$. Preliminary results on the special case
$q_b=0$ are given in \cite{Edalati:2011yv} and we treat the 
case $q_b=2q_f$ in which case a Yukawa coupling is allowed in a companion article \cite{inprogress, {inprogress2}}.


Our main aim is to examine the zero temperature groundstates for
different values of these parameters. This implies that we need to
solve the full equations of motion of this system including the
backreactions of the gauge field and the matter fields on the geometry. 
These equations of motion 
are
\bea
R_{\mu\nu}-\frac{1}{2}g_{\mu\nu}R-\frac{3}{L^2}g_{\mu\nu}&=&\kappa^2\big[T_{\mu\nu}^{\text{gauge}}+T_{\mu\nu}^{\text{fermion}}+T_{\mu\nu}^{\text{boson}}\big],\nonumber\\
{\nabla_\nu}F^{\mu\nu}&=&e^2 \big[J^{\mu}_{\text{boson}}+J^\mu_{\text{fermion}}\big],\nonumber\\
(\nabla^\mu-iq_bA^\mu)(\nabla_\mu-iq_bA_\mu)\phi-\frac{\phi}{2|\phi |}V'(|\phi|)&=&0,\nonumber\\
i(\Gamma^{\mu}\mathcal{D}_{\mu}-m_f)\Psi&=&0,
\eea
where
\bea
T_{\mu\nu}^{\text{gauge}}&=&\frac{1}{e^2}\big(F_{\mu\rho}F_{\nu}^{~\rho}-\frac{1}{4}F^2g_{\mu\nu}\big),\nonumber\\
T_{\mu\nu}^{\text{fermion}}&=&\frac{1}{2}\langle i\bar\Psi\Gamma_{(\mu}\mathcal{D}_{\nu)}\Psi-i\bar\Psi \overleftarrow{\mathcal{D}}_{(\mu}\Gamma_{\nu)}\Psi\rangle,\nonumber\\
T_{\mu\nu}^{\text{boson}}&=&(\partial_\mu+iq_bA_\mu)\phi^*(\partial_\nu-iq_bA_\nu)\phi+(\partial_\mu-iq_bA_\mu)\phi(\partial_\nu+iq_bA_\nu)\phi^*\nonumber\\&&~~-g_{\mu\nu}\big[|(\partial_\alpha-iq_bA_\alpha)\phi|^2+V(|\phi|)\big],\nonumber\\
J^\mu_{\text{fermion}}&=&-q_f\langle\bar{\Psi}\Gamma^\mu\Psi\rangle, \nonumber\\
J^{\mu}_{\text{boson}}&=& -iq_b\big[\phi^*(\partial^\mu-iq_bA^\mu)\phi-\phi(\partial^\mu+iq_bA^\mu)\phi^*\big]
\eea
with $A_{(\mu}B_{\nu)}=\frac{1}{2}(A_\mu B_\nu+A_\nu B_{\mu})$ and $\bar\Psi \overleftarrow{\mathcal{D}}_\mu=\partial_\mu\bar{\Psi} +\frac{1}{4}\omega_{ab \mu}\bar{\Psi} \Gamma^{ab}+iq_fA_\mu\bar{\Psi} $.
The conventions for $\Gamma$-matrices that we use in this paper are
\be
\Gamma^{\underline{t}}=\begin{pmatrix} 
i\sigma^1    & 0\\ 
0 & i\sigma^1  
\end{pmatrix}, ~~~
\Gamma^{\underline{r}}=\begin{pmatrix} 
-\sigma^3    & 0\\ 
0 &-\sigma^3
\end{pmatrix}, ~~~\Gamma^{\underline{x}}=\begin{pmatrix} 
-\sigma^2   & 0\\ 
0 & \sigma^2  
\end{pmatrix},  ~~~\Gamma^{\underline{y}}=\begin{pmatrix} 
0 & \sigma^2  \\
\sigma^2   & 0 
\end{pmatrix}.
\ee

\subsection{Fermions in the fluid approximation} 

The inherent quantum nature of the fermions means the sources for the
backreaction on the geometry and the gauge field are really
expectation values. There are several ways to approximate these
expectation values and incorporate the backreactions of the fermions
in the bulk, which correspond to fermions in different limits: the semi-classical
electron star construction with fermions in the fluid approximation
limit \cite{Hartnoll:2009ns,Hartnoll:2010gu}, the quantum
electron star with fermions treated quantum mechanically
\cite{Sachdev:2011ze,Allais:2012ye,Medvedyeva:2013rpa,Allais:2013lha}.\footnote{At
  finite temperature one can also resort to a single wavefunction
  Dirac hair limit \cite{Cubrovic:2010bf} where the total charge is
  carried by a single radial fermion wavefunction.} We shall stay in
the semi-classical approximation in the following and treat the
fermions in the Thomas-Fermi fluid approximation limit. This is the
well-known Tolman-Oppenheimer-Volkov construction of self-gravitating
stars. For completeness we briefly review it here. Further details are
in \cite{Hartnoll:2010gu,Cubrovic:2011xm}.

The essence of the fluid approximation is adiabaticity of the radial
dependence of the chemical potential, together with a Thomas-Fermi
approximation where we take the number of fermions to infinity while
sending the level spacing to zero. We thus assume that the local
chemical potential varies so slowly $\partial_r \mu_{\text{local}}(r)\ll \mu_{\text{local}}(r)^2$ 
 that we can consider the contribution of fermions as if in a flat
 homogeneous spacetime. 

And in the Thomas-Fermi limit the stress-tensor and charge density of
the fermions take the ideal fluid form
\be
T_{\mu\nu}^{\text{fermion}}=(\rho+p)u_\mu u_\nu+pg_{\mu\nu},~~~J_\mu=-q_f\langle\bar{\Psi}\Gamma_\mu\Psi\rangle= q_f n u_{\mu},
\ee
with $u_t=e_{t\underline{t}}=-\sqrt{-g_{tt}}$ the local Fermi-fluid
velocity. The energy density $\rho$, the pressure $p$ and the number
density $n$ follow directly from an integral over the now
infinitesimally spaced density of states 
\be 
D(\omega)=\frac{1}{\pi^2}\omega\sqrt{\omega^2-m_f^2}.\ee 
\bea\label{rho}
&&\rho=\int_{m_f}^{\mu}\omega D(\omega)d\omega,~~~~~
n=\int_{m_f}^{\mu}
D(\omega)d\omega,~~~~~~
\label{p} p=\mu n-\rho,
\eea 
In the adiabatic approximation the chemical potential $\mu$ is
promoted to local variable, whose evolution is self-consistently
determined from the equations of motion.
A famed characteristic of such self-gravitating semi-classical stars
is that all these fluid parameters will vanish when $\mu\leq m_f$. The
radial value that corresponds to this value of $\mu(r)$ is the edge of
the star
where the full fluid energy, charge density and pressure vanishes.

\bigskip
As we shall search for solutions to the equations of motion we are
implicitly in the semi-classical gravity approximation. For this to be
valid including the backreactions of the matter fields, we need
$\kappa/L \ll 1$ and $\kappa^2 T_{\mu\nu}\sim \mathcal{O}(1)$. This
implies
\begin{itemize}
\item As $\kappa^2 T_{\mu\nu}^{\text{fermion}}\sim \mathcal{O}(1)$, we
  have $\rho, p\sim (\kappa L)^{-2}$.
\item
From (\ref{rho}), we have $\mu\sim (\kappa L)^{-1/2}$ and $m_f\sim
(\kappa L)^{-1/2}$. 
\item
From $\kappa^2 T_{\mu\nu}^{\text{gauge}}\sim \mathcal{O}(1)$, we have
$A_t\sim e L\kappa^{-1}$.
\item 
From $ \kappa^2 T_{\mu\nu}^{\text{boson}}\sim \mathcal{O}(1)$, we have
$\phi\sim 1/\kappa$, $u\sim\kappa^{2}$,
$m_b\sim 1/L$ and $q_b \sim \kappa e^{-1}L^{-1}$. 
\end{itemize}
As $\mu$ is also given by $\mu\sim A_{\underline{t}}$, we must have that
\be
\label{semclass}  e^2 \sim\kappa/L \ll 1.
\ee

It is convenient to rescale all fields and parameters according to their orders in $\kappa$, $e$ and $L$ as follows
\be
p=\frac{1}{\kappa^2L^2}\hat{p},~~~\rho=\frac{1}{\kappa^2L^2}\hat{\rho},~~~n=\frac{1}{e\kappa L^2}\hat{n},
~~A_{\mu}=\frac{e L}{\kappa}\hat{A_{\mu}},\ee
\be u=\kappa^2 \hat{u},~~~m_b=\frac{1}{L}\hat{m}_b,~~~
m_f=\frac{e}{\kappa}\hat{m_f}
,~~~\phi=\frac{1}{\kappa}\hat{\phi},~~~\mu=\frac{e}{\kappa}\hat{\mu},~~~q_b= \frac{\kappa}{eL} \hat{q}_b.
\ee
Thus $V(|\phi|)=\frac{1}{\kappa^2 L^2}\hat{V}(|\hat{\phi}|)$ turns into 
\be \hat{V}(|\hat{\phi}|)=\frac{\hat{u}}{2}(|\hat{\phi}|^2+\frac{\hat{m}_b^2}{\hat{u}})^2-\frac{\hat{m}_b^4}{2\hat{u}},\ee
and the fluid parameters become 
\bea\label{fluidpar}
&&\hat{\rho}=\beta\int_{\hat{m}_f}^{\hat{\mu}}\epsilon^2\sqrt{\epsilon^2-{\hat{m}_f}^2}d\epsilon,~~~~~
\hat{n}=\beta\int_{\hat{m}_f}^{\hat{\mu}}\epsilon\sqrt{\epsilon^2-{\hat{m}_f}^2}d\epsilon,~~~~~
 \hat{p}=\hat{\mu} \hat{n}-\hat{\rho},\eea 
where \be\beta=\frac{e^4 L^2}{\pi^2\kappa^2}\ee
is an $\mathcal{O}(1)$ number.

This rescaling procedure in fact rescales $\kappa$ and $L$ out of the
equations of motion, and only leaves $\beta$ (or alternatively $e$),
$\hat{q}_b$, $\hat{m}_f$, $\hat{m}_b^2$ and $\hat{u}$ as
parameters. Notably no terms in the equations of motions are
irrelevant in this semiclassical limit.
%
%
%

\subsection{Homogenous solutions to the Charged Fluid-Scalar-Gravity system in AdS}

To solve the equations of motion, we make the following homogeneous ansatz of the metric, the gauge field and the matter fields
\be ds^2=L^2\bigg(-f(r)dt^2+g(r)dr^2+r^2(dx^2+dy^2)\bigg),~~~\hat{A}_t=
h(r),~~~\hat{\phi}=
\hat{\phi}(r).\ee
The equations of motion become
\bea\label{hes1}
 \frac{h'^2}{2f}+\frac{f'}{rf}-g\big(3+\hat{p}-\hat{V}\big)+\frac{1}{r^2}-\hat{\phi}'^2-\frac{\hat{q}_b^2gh^2\hat{\phi}^2}{f}=0, &&\\\label{hes2}
  \frac{1}{r}(\frac{f'}{f}+\frac{g'}{g})-g\hat{\mu} \hat{n}-2\hat{\phi}'^2-\frac{2\hat{q}_b^2gh^2\hat{\phi}^2}{f}=0, &&\\\label{hes3}
\label{eomhhh} h''-\frac{h'}{2}(\frac{f'}{f}+\frac{g'}{g}-\frac{4}{r})-\sqrt{f}g\hat{n}-2\hat{q}_b^2g h \hat{\phi}^2=0, &&\\\label{hes4}
\label{eomphi1} \hat{\phi}''+\frac{\hat{\phi}'}{2}\big(\frac{f'}{f}-\frac{g'}{g}+\frac{4}{r}\big)-\frac{1}{2}g\frac{\partial \hat{V}}{\partial{\hat{\phi}}}+\frac{\hat{q}_b^2gh^2\hat{\phi}}{f}=0.
\eea
In addition there is a constraint from the energy momentum conservation
\be
-2\sqrt{f}\hat{n} h'+\hat{\mu}\hat{n} f'+2f\hat{p}'=0.
\ee 
Substituting the fluid expressions (\ref{fluidpar}) into the above
equation, one obtains the relation between the local chemical
potential and the electrostatic potential
\be
\hat{\mu}=\frac{h+C}{\sqrt{f}}.
\ee
We will set the constant $C=0$ to avoid possible singularities.


\bigskip

We shall search for solutions which are asymptotically AdS$_4$. When $r\to \infty$, as $\hat{\mu} \to 0$ all the 
fluid parameters $\hat{\rho}$, $\hat{p}$, $\hat{n}$ vanish. The
asymptotic behavior can therefore be 
analyzed in the framework of Einstein-Maxwell-Scalar gravity. 
This is well known \cite{Hartnoll:2008kx,{Gubser:2009cg},{Horowitz:2009ij}}. 
For a ``light'' scalar field with $-9/4<\hat{m}_b^2<-5/4$ 
 the behavior of the fields near the conformal AdS$_4$ boundary is 
 \bea\label{boundarybehavior}
 f&=&c^2 \big(r^2-\frac{E-(\frac{4}{3}\hat{m}_b^2+2\Delta_1)\Phi_1\Phi_2}{r}\big)+\dots \nonumber\\
 g&=&\frac{1}{r^2}(1-\frac{\Delta_1}{r^{2\Delta_1}}\Phi_1^2+\frac{E-2\Delta_1\Phi_1\Phi_2}{r^3})+\dots \nonumber\\
 h&=&c(\mu-\frac{Q}{r})+\dots\nonumber\\
\hat{\phi} &=&\frac{\Phi_1}{r^{\Delta_{1}}}+\frac{\Phi_2}{r^{\Delta_{2}}}+\dots.
 \eea
Here $\Delta_1, \Delta_2$ are the two roots of the relation $\Delta
(\Delta-3)=\hat{m}_b^2$ and $\Delta_1\leq \Delta_2.$ 
The scalar field has to be normalizable in this background for
self-consistency. The standard boundary condition that ensures this is $\Phi_1=0$ which can be
extended to the full semi-infinite range $\Delta_2>3/2$. An
alternative boundary condition is $\Phi_2=0$ and is only an option
within the range $1/2 < \Delta_1 <3/2$ where $1/2$ is the unitarity
bound. 

\bigskip

The thermodynamic properties of any particular solution are encoded in the value
of its on-shell action. The bulk on shell Lagrangian can be simplified as \cite{Hartnoll:2008kx} 
 \be\sqrt{-g}\mathcal{L}_{\text{on-shell}}=\frac{1}{\kappa^2}\sqrt{-g}g^{xx}R_{xx}=-\frac{L^2}{\kappa^2}\bigg(\sqrt{\frac{f}{g}}r\bigg)'\ee
using the equation of motion for $g_{xx}$.
In addition there is also a boundary term both in the gravity and the
scalar sector. 
 \bea
 \sqrt{-\gamma}\mathcal{L}_{\text
   bnd}&=&\sqrt{-\gamma}\bigg(\frac{1}{\kappa^2}(K-\frac{2}{L})-\frac{\Delta_1}{L}\phi^2\bigg)
 \nonumber\\
\Rightarrow
 \left.\sqrt{-\gamma}\mathcal{L}_{\text
   bnd}\right|_{\text{on-shell}}
 &=&\frac{L^2}{\kappa^2}\sqrt{f}r^2\bigg(\frac{2}{r\sqrt{g}}+\frac{f'}{2f\sqrt{g}}-2-\Delta_1\hat{\phi}^2\bigg) .
 \eea
 Thus the total on-shell action is 
 \be
\frac{\kappa^2}{L^2V}S_{\text{on-shell}}=c \big(E/2-(2\hat{m}_b^2+3)\Phi_1\Phi_2\big).
\ee


The equations of motion have a 
scaling symmetry 
\be\label{scalingsym}
r\to a r, ~~(t,x,y)\to (t,x,y)/a,~~ f\to a^2 f,~~,g\to g/ a^2 ,~~ h\to a h 
\ee 
with an associated Noether current
\be
J_N=\frac{-2r^2hh'+r^2 f'-2rf}{\sqrt{fg}}
\ee
This
can be used to derive the following useful 
identity: 
as $\partial_r J_N=0$, we have $J_N(r=\infty)=J_N(r=0)$ and this gives 
\be\label{notheridentity}
3E-2\mu Q-(4\hat{m}_b^2+6\Delta_1)\Phi_1\Phi_2=0
\ee
at zero temperature.\footnote{At finite temperature, as $J_N(r_{\text{hor}})\neq 0$ there will be an extra term in the eqn. (\ref{notheridentity}).}
This relation is extremal to check the consistency of the numerical
solutions we shall derive below. In addition it helps expose the
underlying thermodynamics in the gravity system. With the help of this
identity the free energy -- minus the on-shell effective action --- 
can be rewritten in the standard zero-temperature equilibrium relation
\be F/V=-\frac{\kappa^2}{cL^2V}S_{\text{on-shell}}=E-\mu Q. 
\ee

Let us briefly discuss the total charge density on the right-hand-side
of the above equation. In any multi component system, it is the sum of
individual contributions.
In the system we study here with both fermions and bosons, the total charge density of the dual field theory is 
composed out of the bosonic charge density and the fermionic charge
density. We can see this explicitly. The total boundary charge density can be read from the asymptotic behavior of the Maxwell field \be\rho_{\text{boundary}}=\sqrt{-g} F^{tr}|_{r=\infty}=\frac{h'r^2}{\sqrt{fg}}\bigg{|}_{r=\infty}=Q.\ee

Inspecting the equations of motion of this system, we see we can
rewrite (\ref{eomhhh}) as \be Q=Q_b+Q_f,\ee where the bulk charged
densities integrated along the radial direction are directly
recognized  as
\be Q_b=\int_{0}^{\infty} dr \sqrt{-g} \frac{2 \hat{q}_b^2 h\hat{\phi}^2}{f}\ee and \be Q_f=\int_{0}^{\infty} dr\sqrt{-g} \frac{\hat{n}}{\sqrt{f}}.\ee 

\bigskip

In the next section we will search for solutions of this system by finding the near horizon solutions first and then integrate to the asymptotically AdS$_4$ boundary with normalizable scalar boundary conditions.

\section{Zero temperature solutions} 
\label{zerotsol}
   
We shall aim to determine the most stable homogeneous ground states at
zero temperature for different parameter regions of
($\hat{m_b}$,$\hat{m_f}$,$\hat{q_b}$), holding $\beta$ (or
equivalently $q_f$) and $\hat{u}$ fixed to simplify the system.  We
shall find that in addition to the three known types of zero
temperature solutions in this system: the  AdS Reissner Nordstr\"om
(RN) black hole
($T_{\mu\nu}^{\text{boson}}=T_{\mu\nu}^{\text{fermion}}=0$), the
holographic superconductor solution ($T_{\mu\nu}^{\text{boson}} \neq
0$,~$T_{\mu\nu}^{\text{fermion}}=0$), and the electron star solution
($T_{\mu\nu}^{\text{boson}}=0$,~$T_{\mu\nu}^{\text{fermion}}\neq 0$),
there is in addition a new kind of hairy electron star solution for
which $T_{\mu\nu}^{\text{boson}} \neq 0$ and
$T_{\mu\nu}^{\text{fermion}}\neq 0$. 

Let us briefly summarize the three known solutions:
\begin{itemize}
\item The Reissner Nordstr\"om black hole (RN) is the solution to this system when no scalar field or fermionic fluid is excited. The solution is
 \be\label{RNAdS}
f(r)=\frac{1}{g(r)}=r^2\bigg(1+\frac{Q^2}{r^4}-\frac{1}{r^3}(r_0^3+\frac{Q^2}{r_0})\bigg), ~~~h=\frac{\sqrt{2}Q}{r_0}\big(1-\frac{r_0}{r}\big),
 \ee where $r_0$ is the horizon, $Q$ is the charge and at zero
 temperature $Q=\sqrt{3}r_0^2$. The zero-temperature near horizon geometry is AdS$_2$ $\times$ R$^2$ with the AdS$_2$ radius $L_2=L/\sqrt{6}.$ In grand canonical enssemble, the free energy is $F/\mu^{3}\simeq-0.136$. This solution corresponds to the disordered phase of the boundary field theory.

\item The holographic superconductor (HS) solution \cite{Gubser:2009cg, {Horowitz:2009ij}} is the solution to this system where only normalizable bosons are excited. For nonzero $\hat{u}$, the near horizon geometry of zero temperature holographic superconductor solutions is Lifshitz geometry, {\it i.e. } when $r\to 0$, we have\footnote{For $\hat{q}_b$=0 case \cite{Iqbal:2010eh} one should use the near horizon ansatz AdS$_2$$\times$ R$^2$.} 
\be\label{lifnh}
f= r^{2z},~~~g= \frac{g_0}{r^2},~~~h=h_0 r^z,~~~\hat{\phi}=\hat{\phi}_0. 
\ee The constants $(g_0,h_0,z,\hat{\phi}_0)$ are determined by the
parameters of the system
$(\hat{m}_b,\hat{u},\hat{q_b})$. (See eqn. (\ref{nhlif})). 
This solution is dual to a superconducting phase at the boundary.

\item The electron star (ES) solution \cite{Hartnoll:2010gu} is the
  solution when only fermions are excited, approximated by a
  fermion-fluid description. The near horizon geometry is also
  Lifshitz like as in (\ref{lifnh}) but with
  $\hat{\phi}_0=0$. (See eqn. (\ref{nhlifes})). 
  This is
  dual to a Fermi liquid with multiple Fermi surfaces at the boundary \cite{Hartnoll:2011dm, {Iqbal:2011in}, {Cubrovic:2011xm}}.

\end{itemize}

\subsection{IR Stability analysis}

If the rule of thumb that bosons always win is correct, then there is
a quite direct way to test this with a simple stability
analysis. Starting from the electron star solution we add the scalar
field as a probe and check for whether it becomes unstable in the near-horizon region. 
We know that the holographic superconductor background is more stable
than the AdS RN background when the mass of the bosons is below the BF
bound in the standard quantization. 
The BF bound is essentially the effective mass of the scalar field in
the near-horizon region. Consider then the electron star background
instead of AdS-RN. In the presence of fermions, this 
electron star is always more stable than the AdS-RN background at
zero temperature as long as $\hat{m}_f<1$. As the fermions and bosons
do not have a direct interaction, the relevancy for the stability
analysis is that the near-horizon ES background is now charged
Lifshitz. The scalar-field equation of motion in this background
is
\be
\hat{\phi}''+\frac{3+z}{r}\hat{\phi}'-\frac{g_0}{r^2}(\hat{m}_b^2-h_0^2\hat{q}_b^2)\hat{\phi}-\frac{\hat{u}g_0}{r^2}\hat{\phi}^3=0.
\ee 
 Thus we see that the BF bound instability condition for charged
bosons in the electron star background is 
\be\label{estohesbfbound}
\hat{m}_b^2-h_0^2\hat{q}_b^2 \leq -\frac{(2+z)^2}{4g_0}
\ee 
for the standard quantization of the scalar field. Substituting the
relation between the Lifshitz parameters and the fermion mass (recall
that the charge is fixed to unity) one sees that for each $\hat{m}_f$ there
is indeed a critical value of $\hat{m}_b^2$ for which the scalar condenses.

\begin{figure}[h]
\begin{center}
\begin{tabular}{cc}
\includegraphics[width=0.98\textwidth]{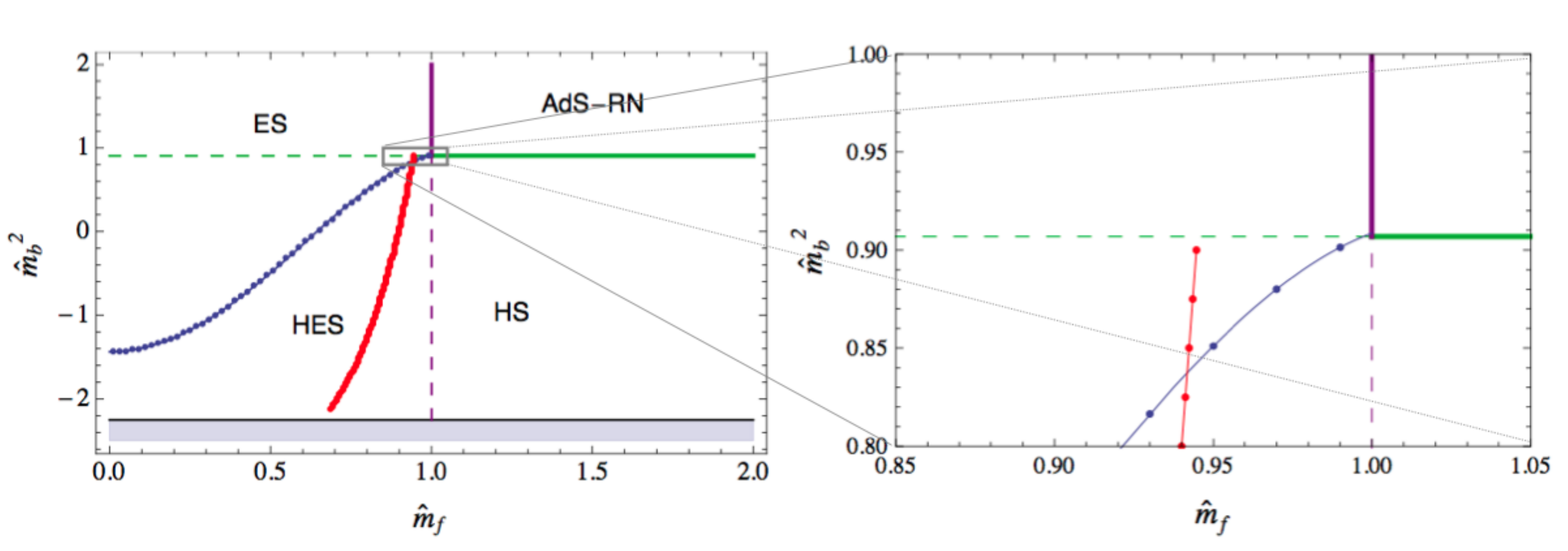}
\end{tabular}
\caption{\small The instability curves for fermions and standard quantization scalars in the 
$\hat{m}_b^2$-$\hat{m}_f$ plane. For continuous phase transitions this should be the phase diagram. 
Here we fix: $(\hat{q}_b,\hat{u},\beta)\simeq (1.55,6,19.951)$. 
The green solid line gives the phase transition between AdS-RN and HS and the purple solid line is the phase transition between AdS-RN and ES. For illustration the green and purple dashed line are the extensions of the corresponding solid line. 
{\it Left:} blue dotted line characterizes the instability for scalar in ES.  
When $\hat{q}_b$ changes slightly, the phase diagram will also change quantitatively, but the qualitative feature stays the same except at $\hat{q}_b=0$. 
The red dotted line is the fermionic BF bound in the near horizon HS region. This gives rise to a puzzle which we shall resolve by computing the exact phase diagram. The puzzle is highlighted on the {\it right}: 
a magnification of the gray-box region in the left figure where the various phases meet. It is seen that the red dotted line (the BF bound) does not cross the other three boundaries at the critical point, whereas continuity between HES and ES would argue that it should. }
\label{figBF-bounds}
\end{center}
\end{figure}

If this condensation indeed signals a transition to the pure bosonic
groundstate, the holographic
superconductor, then one should simultaneously see that the
holographic superconductor in the presence of fermions has an
instability at the same locus in the phase diagram.
From the near horizon solution of holographic superconductor (\ref{nhlif}) we know that the local chemical potential at the horizon is 
\be
\hat{\mu}_{\text{loc}}=\frac{h}{\sqrt{f}}\Big{\arrowvert}_{r_{\text{hor}}}=h_0=\sqrt{1-\frac{1}{z}}.
\ee
When $\hat{m}_f$ is less than this number the system can support a Fermi liquid. 
We therefore see that the instability condition for fermions in the near horizon region is \be\label{hesversushs}
\hat{m}_f<\sqrt{1-\frac{1}{z}}.
\ee 
Substituting for the Lifshitz parameters their expression in terms of
the scalar properties, we can draw both instability curves in a phase
diagram (Fig. \ref{figBF-bounds}) as a function of $(\hat{m}_f, \hat{m}_b^2)$. We immediately see that the two curves do not
coincide, but that there ought to exist an intermediate phase where
{\em both} the fermions and bosons are excited, in other words a hairy
electron star. 
This is a new state and it corresponds to a phase which has both superconductivity and multiple Fermi surfaces. However, these fermions are not those which form the Cooper pairs responsible for this superconductivity because there is no direct BCS type interaction between them and the charges are not related. The system with BCS interactions will be studied in \cite{inprogress}.

Now that we know this solution has to exist, we will construct it
explicitly.
Before we do so, however, note that the instability analysis reveals a
curious aspect. Zooming in on the location where the phase boundaries
intersect, we see that the fermion instability curve in the HS phase does not smoothly
transition into the ES-AdS-RN phase boundary at the critical values of
$\hat{m}_b^2$.
 This is puzzling and could mean various things, such as a
missed degree of freedom.\footnote{Note that the Gibbs rule forbidding a
quadruple point does not apply as we have multicomponent system.} We will see that the explicit solution will
provide the explanation.



\subsection{The new Hairy Electron Star solution}

To obtain the hairy electron star solution we follow the same
procedure as used for the holographic superconductor and the electron
star. We will show that the near horizon geometry in this case is also
Lifshitz geometry. Then the full asymptotically AdS solutions can be
obtained by turning on the irrelevant deformations near the Lifshitz
fixed points (\ref{lifnh}). 

To illustrate the commonality of the new solution with the other
near-horizon Lifshitz solutions, i.e. the holographic superconductor and the
electron star, let us first briefly review the details of these solutions 
to continue with the construction of the hairy
electron star solution. In the subsequent section we will analyze the
thermodynamics of all solutions leading to the phase diagram.

\bigskip
{\bf The Holographic Superconductor: }
In the absence of fermions the equations of motion simplify to 
\cite{Gubser:2009cg,{Hartnoll:2008kx},{Horowitz:2009ij}}
\bea\label{eomhs1}
 \frac{h'^2}{2f}+\frac{f'}{rf}-g\big(3-\hat{V}\big)+\frac{1}{r^2}-\hat{\phi}'^2-\frac{\hat{q}_b^2gh^2\hat{\phi}^2}{f}=0, &&\\\label{eomhs2}
  \frac{1}{r}(\frac{f'}{f}+\frac{g'}{g})-2\hat{\phi}'^2-\frac{2\hat{q}_b^2gh^2\hat{\phi}^2}{f}=0, &&\\\label{eomhs3}
 h''-\frac{h'}{2}(\frac{f'}{f}+\frac{g'}{g}-\frac{4}{r})-2\hat{q}_b^2g h \hat{\phi}^2=0, &&\\\label{eomhs4}
 \hat{\phi}''+\frac{\hat{\phi}'}{2}\big(\frac{f'}{f}-\frac{g'}{g}+\frac{4}{r}\big)-\frac{1}{2}g\frac{\partial \hat{V}}{\partial{\hat{\phi}}}+\frac{\hat{q}_b^2gh^2\hat{\phi}}{f}=0.
\eea
With the ansatz (\ref{lifnh}), this system has a Lifshitz scaling solution\footnote{There is another possible solution with AdS$_4$ near horizon geometry: $g_0=1, z=1,\hat{\phi}_0=0$, which usually has a higher free energy and we will not consider it here.} 
\bea\label{nhlif}
h_0^2&=&\frac{z-1}{z},\nonumber\\
\hat{\phi}_0^2&=&\frac{6z}{\hat{m}_b^2 z+\hat{q}_b^2(3+2 z+z^2)},\nonumber\\
g_0&=&\frac{z}{\hat{q}_b^2\hat{\phi}_0^2}=\frac{\hat{m}_b^2 z+\hat{q}_b^2(3+2 z+z^2)}{6\hat{q}_b^2},\nonumber\\
\hat{u}&=&-\frac{\hat{m}_b^4z^2+\hat{m}_b^2\hat{q}_b^2z(4+z+z^2)-\hat{q}_b^4(-3+z+z^2+z^3)}{6z^2}
\eea
There is a natural constraint $z\geq 1$ to make sure that $h_0$ is a real constant. For fixed values of $\hat{u}$ and $\hat{m}_b$, $z$ decreases when $\hat{q}_b$ increases and the condition $z\geq 1$ gives a constraint on $\hat{q}_b$ that $\hat{q}_b$ has a maximum value $\hat{q}_{b,\text{max}}(\hat{u},\hat{m}_b)$ at which $z=1$.

The holographic superconductor is the domain wall solution that
interpolates between asymptotic $AdS_4$ and this Lifshitz solution in
the interior. Integrating outwards,
we need to consider the following irrelevant perturbation from the near horizon Lifshitz solution to flow to $AdS_4$ on the boundary \cite{Gubser:2009cg}
\bea
f&=& r^{2z}\big(1+f_1 r^{\alpha_1}+f_2 r^{\alpha_2}\big)+\dots,\nonumber\\
g&=& \frac{g_0}{r^2}\big(1+g_1 r^{\alpha_1}+g_2 r^{\alpha_2}\big)+\dots,\nonumber\\
h&=&h_0 r^z\big(1+h_1 r^{\alpha_1}+h_2 r^{\alpha_2}\big)+\dots,\nonumber\\
\hat{\phi}&=&\hat{\phi}_0\big(1+\hat{\phi}_1 r^{\alpha_1}+\hat{\phi}_2 r^{\alpha_2}\big)+\dots,
\eea  where $\alpha_1>\alpha_2>0$ are the roots of the sextic equation for $\alpha$ 
\be
\alpha(\alpha+2+z)\big(\alpha^4+(4+2z)\alpha^3+C_2\alpha^2+C_1\alpha+C_0\big)=0
\ee
where 
\bea
C_0&=&-6\hat{m}_b^2+6\hat{q}_b^2-\frac{6\hat{q}_b^2}{z}+\frac{20\hat{m}_b^2z}{3}-\frac{4\hat{m}_b^4z}{3\hat{q}_b^2}-\frac{4\hat{q}_b^2z}{3}-\frac{4\hat{m}_b^2z^2}{3}
+\frac{2\hat{m}_b^4z^2}{\hat{q}_b^2}\nonumber\\&&+\frac{4\hat{q}_b^2z^2}{3}+\frac{4\hat{m}_b^2z^3}{3}-\frac{2\hat{m}_b^4z^3}{3\hat{q}_b^2}-\frac{2\hat{q}_b^2z^3}{3}-\frac{2\hat{m}_b^2z^4}{3}+
\frac{2\hat{q}_b^2z^4}{3}
\nonumber\\
C_1&=&-8+\frac{8\hat{m}_b^2}{3}+\frac{\hat{q}_b^2}{3}+\frac{2\hat{q}_b^2}{z}+8z+2\hat{m}_b^2z+\frac{2\hat{m}_b^4z}{3\hat{q}_b^2}-\hat{q}_b^2z+2z^2+\hat{m}_b^2z^2\nonumber\\&&
+\frac{\hat{m}_b^4z^2}{3\hat{q}_b^2}-\hat{q}_b^2z^2-2z^3+\frac{\hat{m}_b^2z^3}{3}-\frac{\hat{q}_b^2z^3}{3}
\nonumber\\
C_2&=&\frac{4\hat{m}_b^2}{3}-\frac{\hat{q}_b^2}{3}+\frac{\hat{q}_b^2}{z}+10z+\frac{\hat{m}_b^2z}{3}+\frac{\hat{m}_b^4z}{3\hat{q}_b^2}-\frac{\hat{q}_b^2z}{3}-z^2+\frac{\hat{m}_b^2z^2}{3}-\frac{\hat{q}_b^2z^2}{3}.
\eea
There are two independent perturbations because the equation of motion for the scalar field is 
second order and we need an additional degree of freedom at the horizon to
satisfy the normalizability boundary condition at the boundary.  
It is important to note that when both $\alpha_1$ and $\alpha_2$ are real only the relative ratio of $f_1/f_2$ is nontrivial since 
we can rescale $f_1$ or $f_2$ to $1$ by rescaling the coordinate $r$ (the sign of $f_1$ or $f_2$ still matters).  The universal solution $\alpha=-2-z$ is related to making the Lifshitz background nonextremal.  From the other four solutions, we pick the irrelevant ones,  {\it i.e.} $\text{Re}\alpha_1, \text{Re}\alpha_2>0$ in order to construct an upwards flow from Lifshitz to the  conformal AdS$_4$ boundary. 

\begin{figure}[h]
\begin{center}
\begin{tabular}{cc}
\includegraphics[width=0.49\textwidth]{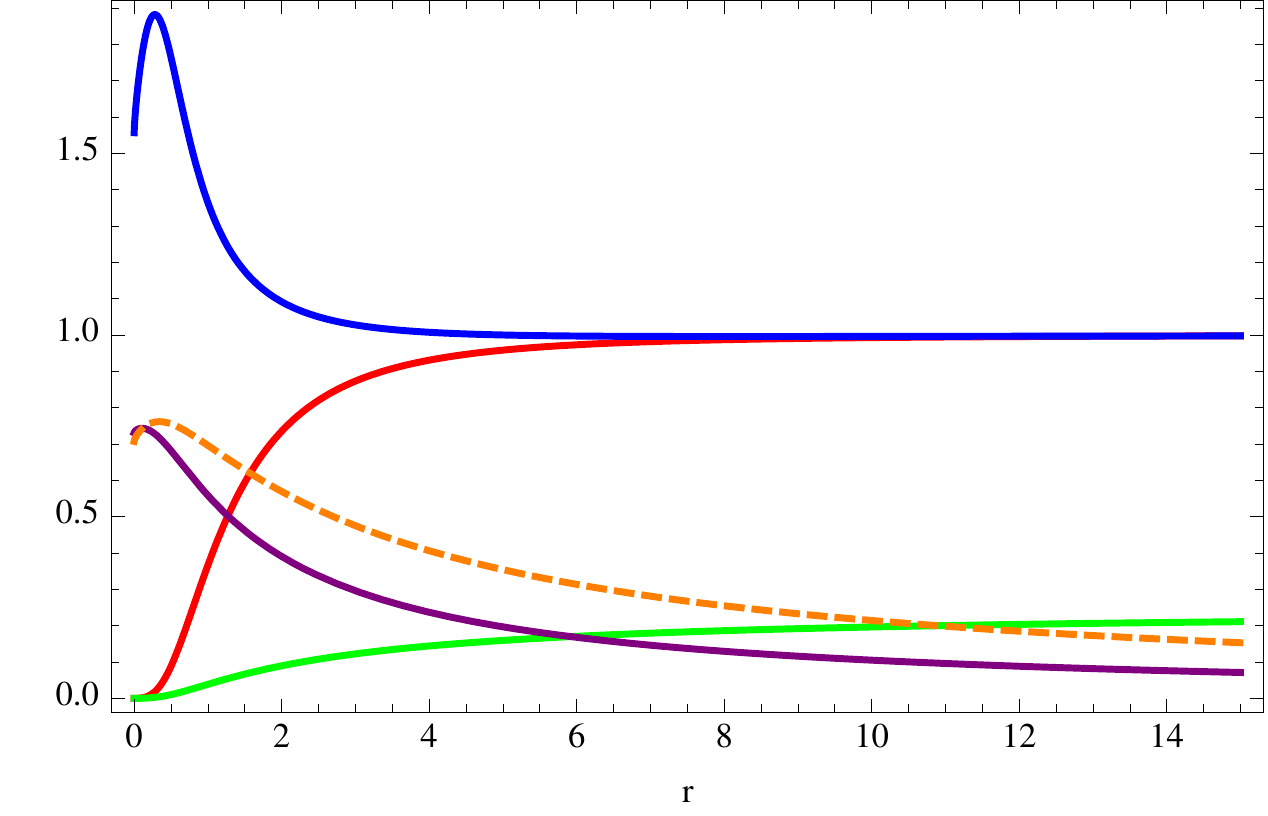}
\includegraphics[width=0.46\textwidth]{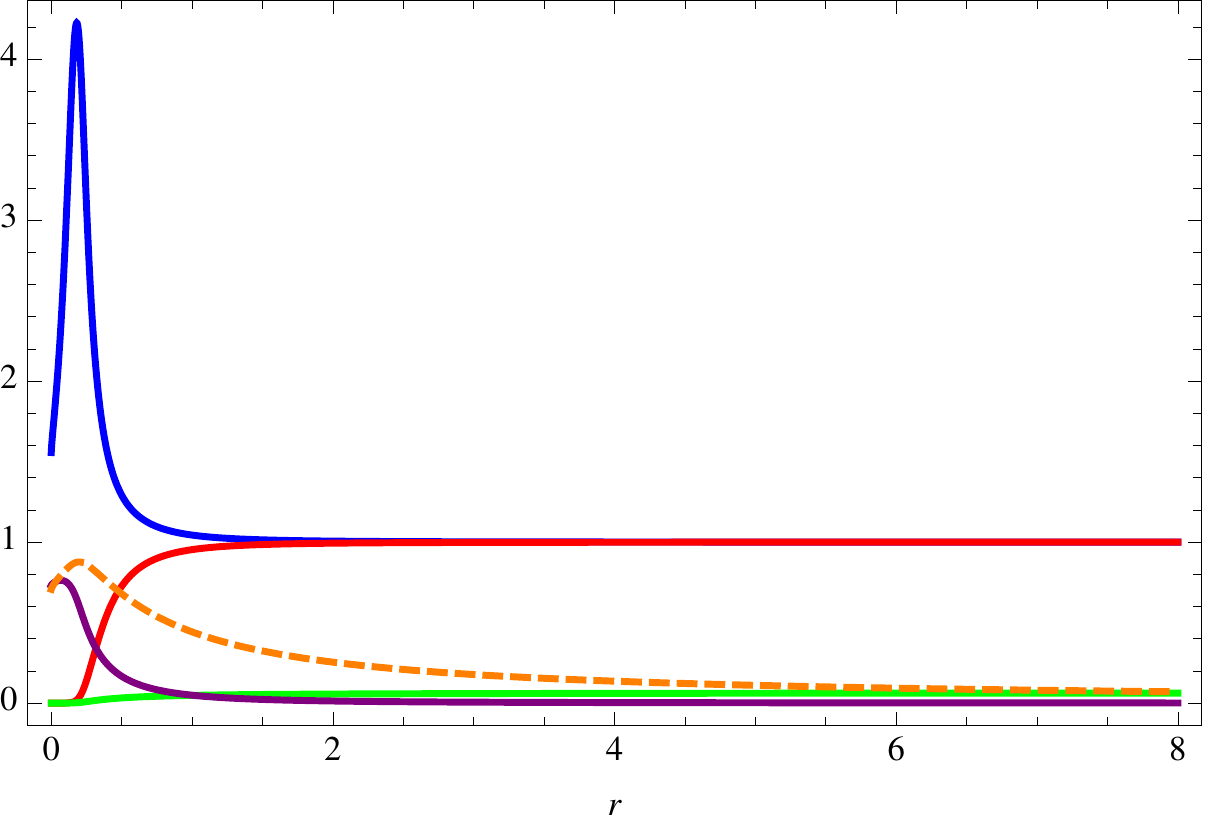}
\end{tabular}
\caption{\small Background of holographic superconductor for the example mentioned in the text above: $f/c^2r^2 $(red), $g r^2$(blue), $h/c\mu$ (green), $\hat{\phi}$ (purple), $\mu_{\text{loc}}$(orange). {\it Left:} alternative quantization case; {\it Right:} standard quantization case. We can see that in both these two cases the local chemical potential has a maximum at an intermediate $r$.}\label{bghs}
\end{center}
\end{figure}

As an example, consider $(\hat{m}_b^2,\hat{q}_b,
\hat{u})=(-2,1.55,6)$. This gives $(z, g_0, h_0,\hat{\phi}_0)\simeq(2,
1.556, 0.707, 0.73)$. Then $(g_1,h_1,\hat{\phi}_1)\simeq(0.336,0.619,
0.129)f_1$, $(g_2,h_2,\hat{\phi}_2)\simeq(0.191,0.577, 0.044)f_2$ and
there are five possible values of $\alpha$. Two of them are real
positive numbers: $(\alpha_1,\alpha_2)\simeq (1.245,
0.728)$
and we select these.
Then, integrating outwards for $(f_1,f_2)=
(3.845,-1)$ one finds a normalizable solution for the scalar field
with $\Phi_2=0$ near the conformal boundary. In this particular 
case, we have $(c,\Phi_1, \mu, Q, E)\simeq
(4.095, 1.085,2.654, 5.896, 10.433)$.  
  In the grand canonical ensemble 
  we have
  $F/\mu^{3}\simeq-0.279.$. For $(f_1,f_2)\simeq (10, 2.16)$ one
obtains the normalizable solution for the standard quantization case
$\Phi_1=0$. Here $F/\mu^3\simeq -0.148.$ In both the standard and
alternative quantization case the qualitative behaviors  of the background
fields profile are the same, 
 see Fig \ref{bghs}. In addition to the fields of the solution, we have drawn
 one other value, the local chemical potential $\mu_{\text{loc}} =
 h/\sqrt{f}$ experienced by fermions. The immediately notably feature
 is that it rises first as we move outward before it decreases. We
 will discuss the importance of this when we construct the electron
 star solution.


{\bf The Electron Star:}
Without any bosonic excitations the equations of motion simplify to \cite{Hartnoll:2010gu,  {deBoer:2009wk},{Arsiwalla:2010bt}}
 \bea\label{eomes1}
 \frac{h'^2}{2f}+\frac{f'}{rf}-g\big(3+\hat{p}\big)+\frac{1}{r^2}=0, &&\\\label{eomes2}
  \frac{1}{r}(\frac{f'}{f}+\frac{g'}{g})-g\hat{\mu} \hat{n}=0, &&\\\label{eomes3}
\label{eomh1} h''-\frac{h'}{2}(\frac{f'}{f}+\frac{g'}{g}-\frac{4}{r})-\sqrt{f}g\hat{n}=0.
\eea
With the ansatz (\ref{lifnh}) and $\hat{\phi}_0=0$, these equations
have the Lifshitz scaling solution with
\be\label{nhlifes}
h_0^2=\frac{z-1}{z},~~~
g_0^2=\frac{36 z^4(z-1)}{((1-\hat{m}_f^2)z-1)^3\beta^2}
\ee
and 
\be
12(2+4z+h_0^2 z^2)+g_0 (-72-2h_0^3\sqrt{h_0^2-m_f^2}\beta+5h_0 m_f^2\sqrt{h_0^2-m_f^2}\beta)+3g_0 m_f^4\beta\log{\frac{m_f}{h_0+\sqrt{h_0^2-m_f^2}}}=0.
\ee
We also have $z>1$ in this case. To flow to $AdS_4$ at the boundary,
we again consider the irrelevant perturbations from the near horizon
Lifshitz solution
\bea\label{nhes}
f&=& r^{2z}\big(1+f_1 r^{\alpha}\big)+\dots,\nonumber\\
g&=& \frac{g_0}{r^2}\big(1+g_1 r^{\alpha}\big)+\dots,\nonumber\\
h&=&h_0 r^z\big(1+h_1 r^{\alpha}\big)+\dots.
\eea
As before, $f_1$ can be rescaled to be $1$ or $-1$.
A new feature is that there will be a specific edge of the star $r_s$
where the local chemical potential equal the mass
$\frac{h(r_s)}{\sqrt{f(r_s)}}=\hat{m}_f.$. Beyond this value, no
fermion fluid can be supported and the solution is matched on that of
a standard AdS Reissner Nordstrom black hole, 
\be
f=c^2 \bigg(r^2-\frac{E}{r}+\frac{Q^2}{2r^2}\bigg),~~g=\frac{c^2}{f},~~h=c\bigg(\mu-\frac{Q}{r}\bigg).
\ee

\begin{figure}[h]
\begin{center}
\begin{tabular}{cc}
\includegraphics[width=0.5\textwidth]{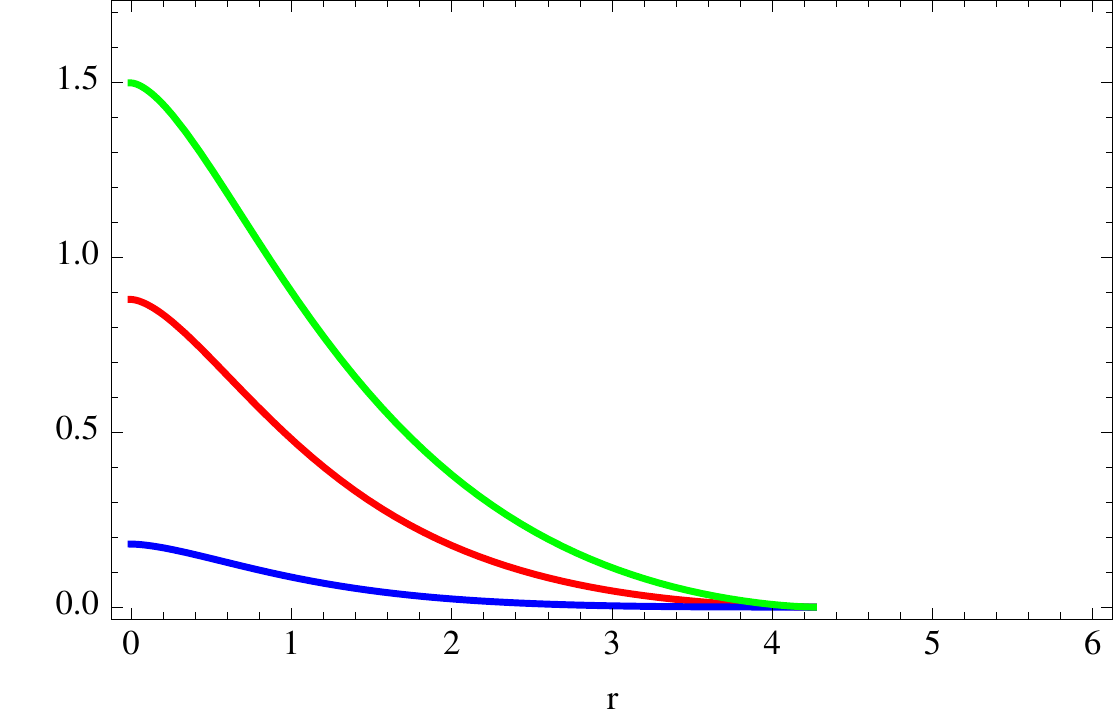}
\includegraphics[width=0.5\textwidth]{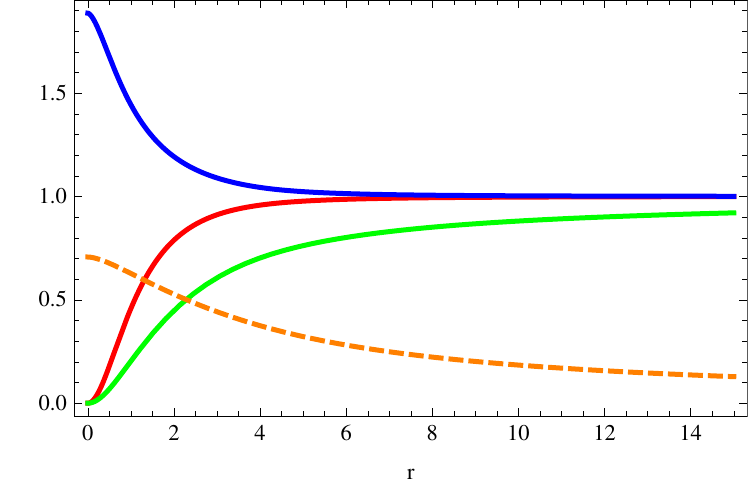}
\end{tabular}
\caption{\small The background solution of an electron star for the specific example mentioned in the text above.  {\it Left:} fluid profiles as functions of the radial coordinate $r$: $\hat{\rho}$ (red), $\hat{n}$ (green), $\hat{p}$ (blue). {\it Right:}  metric background of the electron star: $f/c^2r^2 $(red), $g r^2$(blue), $h/c \mu$ (green), $\mu_{\text{loc}}$(orange).}
\label{bges}
\end{center}
\end{figure}

As an example we consider $(\hat{m}_f, z, \beta) \simeq (0.36,2,19.951)$. 
In this case $(g_0, h_0)\simeq (1.887, 0.707).$ We have four solutions for
 $\alpha.$ The irrelevant one is $\alpha\simeq 1.626$ and $(g_1, h_1)\simeq (0.446, 0.645)f_1$. We choose $f_1=-1$ and we have $(r_s, c, \mu, Q,  E)\simeq (4.256, 1.021, 2.088, 2.483, 3.457)$. In the canonical ensemble, the free energy is 
 $F/\mu^{3}\simeq-0.190.$ 
The geometry background for this parameter set is plotted in
Fig. \ref{bges}.
Note that in this zero-temperature electron star the local chemical
potential is monotonically decreasing. Correspondingly there is a
dense core of the star at $r=0$ which dilutes as one moves outward
(Left figure in Fig \ref{bges}.)

\bigskip

\begin{figure}[h]
\begin{center}
\begin{tabular}{cc}
\includegraphics[width=0.3\textwidth]{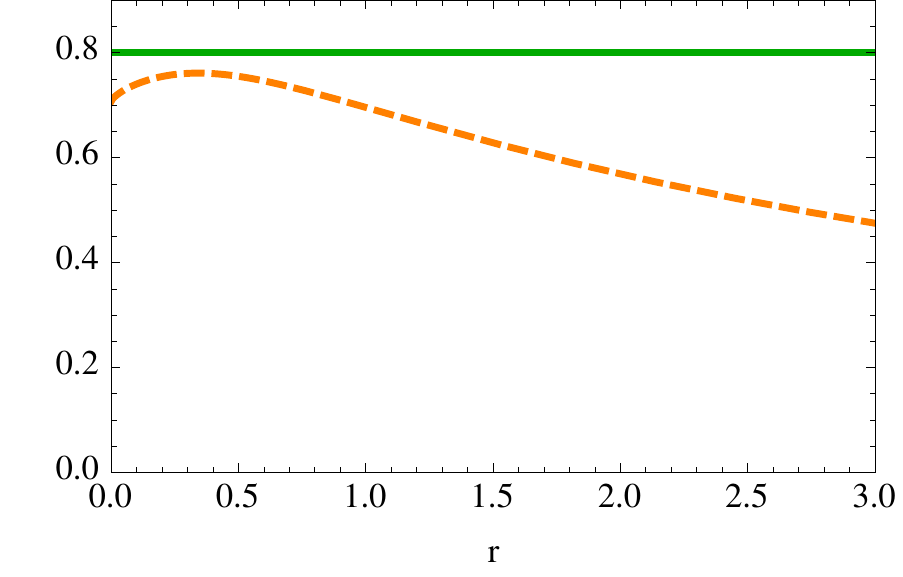}
\includegraphics[width=0.3\textwidth]{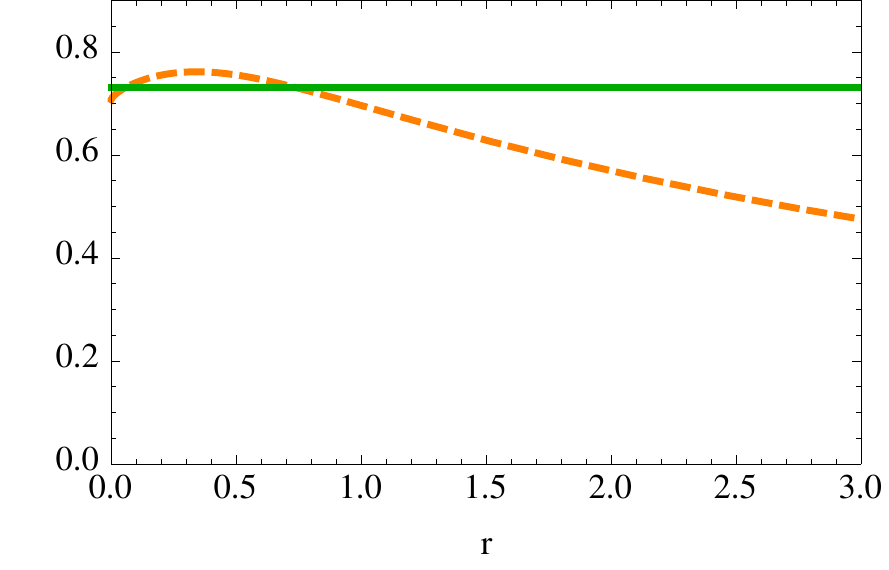}
\includegraphics[width=0.3\textwidth]{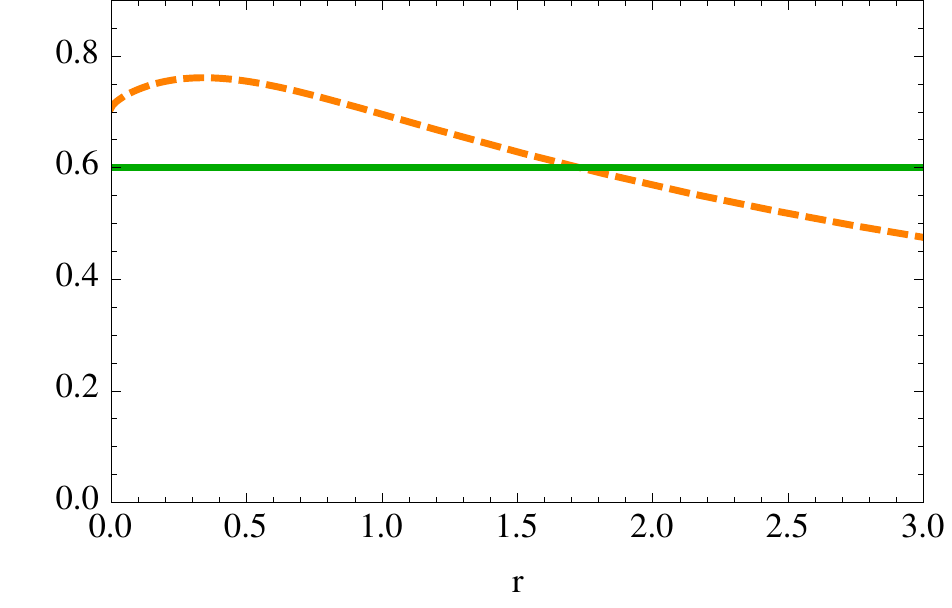}
\end{tabular}
\caption{\small The orange dashed line is the local chemical potential for HS background. When we decrease $\hat{m}_f$ (green line) from left to right, we have HS, two-edge HES and one-edge HES solutions respectively. }
\label{appeares}
\end{center}
\end{figure}

{\bf The Hairy Electron Star:}
Both fermions and bosons materialize and it is now required to
solve the full set of equations of motion. We find it to be useful  to consider
the hairy star solution as an electron star solution that lives on a 
holographic superconducting background. As the fermions and bosons have
no direct interaction this captures the essence of the nature of the full
solution. By tuning the fermion mass downward from a  high value
to below the local chemical potential, the circumstances are created to form a Fermi-fluid. 
However, different from the pure electron star case in
the zero temperature holographic superconductor background the profile of the local chemical potential 
is not monotonic. This type of profile is also known from finite
temperature electron star solutions \cite{Puletti:2010de,{Hartnoll:2010ik}}. 
It implies that there are two possible kinds of hairy electron star backgrounds depending on 
the value of $\hat{m}_f$: If $\hat{m}_f$ is very low, the fermi fluid
can continuously exist from the interior to the outer edge. If,
however, $\hat{m}_f$ is just below the critical value where a fermi
fluid can exist, this fermi fluid has both an inner and an outer edge. This key insight is illustrated in Fig. \ref{appeares}.  
\\

{\bf CASE I: Hairy ES with one edge:}
For a fixed set of parameters, when we tune $\hat{m}_f$ to be small
enough, there can always be fermonic excitations near the horizon.  
The
near horizon parameters $\phi_0$, $g_0$, $h_0$, $z$ again are
determined by the mass and charge parameters, and we do not bother to
write the complicated expression out here.

As an example of hairy electron star solutions with only one edge, we
consider $(\hat{m}_b^2,\hat{q},  \hat{u}, \hat{m}_f
,\beta)\simeq(-2,1.55, 6, 0.36, 19.951)$.\footnote{
For numerical convenience, we only consider systems with $\hat{m}_b^2 =-2$ to obtain the free energies. 
} 
For the Lifshitz parameters
we have $(z, g_0, h_0,\hat{\phi}_0)\simeq(1.284, 1.059, 0.471, 0.650)$. 
Inside the star, the system is described by eqn. (\ref{hes1} - \ref{hes4}).

Because there are bosons present, let us again consider irrelevant perturbations with two terms in order for the solution to flow to $AdS_4$ with normalizable bosons at the boundary:\footnote{\label{foot3} The $\hat{q}_b=0$ case was considered in \cite{Edalati:2011yv}. A crucial difference is the ansatz of the perturbation of the scalar field in the near horizon region. In $\hat{q}_b=0$ case, from the EOM (\ref{eomphi1}) for the scalar field, the metric and the gauge field fluctuations do not affect the scalar field at the first order of perturbation since $V'(\hat{\phi}_0)=0$.}
\bea
f&=& r^{2z}\big(1+f_1 r^{\alpha_1}+f_2 r^{\alpha_2}\big)+\dots,\nonumber\\
g&=& \frac{g_0}{r^2}\big(1+g_1 r^{\alpha_1}+g_2 r^{\alpha_2}\big)+\dots,\nonumber\\
h&=&h_0 r^z\big(1+h_1 r^{\alpha_1}+h_2 r^{\alpha_2}\big)+\dots,\nonumber\\
\hat{\phi}&=&\hat{\phi}_0\big(1+\hat{\phi}_1 r^{\alpha_1}+\hat{\phi}_2 r^{\alpha_2}\big)+\dots. 
\eea
The order of the equation for $\alpha$ is six and again we have the two universal solutions $\alpha_1=0, \alpha_2=-2-z$ which do not depend on the parameters we choose. In this example above, the solutions of $\alpha$ yielding irrelevant perturbations are $\alpha=0.884\pm 0.228 i$. Usually a relevant and complex deformation $\alpha$ implies an instability   \cite{Hartnoll:2011pp,{Gouteraux:2012yr},{Iqbal:2011aj}}. However, it is not yet clear that an irrelevant complex deformation means an instability \cite{Gubser:2009cg}. 
The one notable effect of the complex scaling dimension is that the approach to the Lifshitz fixed point is oscillatory although this is in a very small region. 
We will not discuss this possible instability issue or its relation on the oscillatory approach 
and we will assume that the absence of a relevant deformation
indicates that it is a consistent and stable solution. 
When $\alpha_1$ and $\alpha_2$ are conjugate complex numbers, the scaling symmetry can be used to fix $f_1=f_2$ to be real.

In this specific case we have $(g_1,h_1,\hat{\phi}_1)\simeq(0.274+0.078i, 0.958+0.086i, 0.180+0.176i)f_1$ 
and $(g_2,h_2,\hat{\phi}_2)\simeq(0.274-0.078i, 0.958-0.086i, 0.180-0.176i)f_2$. Note that the functions $f,~g$, and $h, \phi$ are still real. For practical reason here we chose a conjugate $f_{1,2}\simeq -1\pm 6.309i.$ 
The edge of the star $r_s$ where the fermi fluid can no longer be
supported is again defined by the equality of the fermion mass with the
local chemical potential
\be\label{staredge}
\hat{m}_f=\frac{h(r_s)}{\sqrt{f(r_s)}},
\ee
For the numerical values we find $r_s\simeq 0.461.$

\begin{figure}[h]
\begin{center}
\begin{tabular}{cc}
\includegraphics[width=0.5\textwidth]{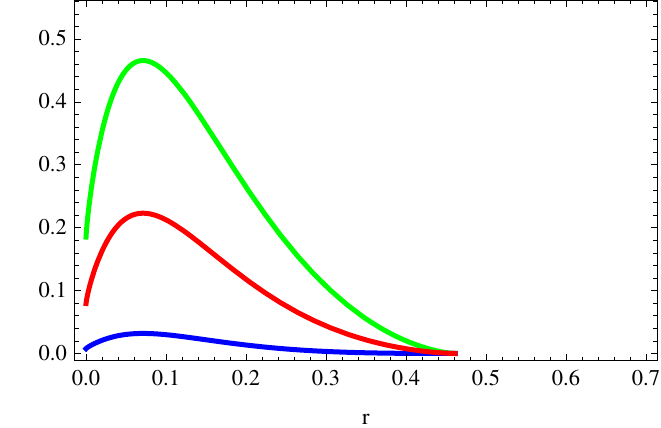}
\includegraphics[width=0.5\textwidth]{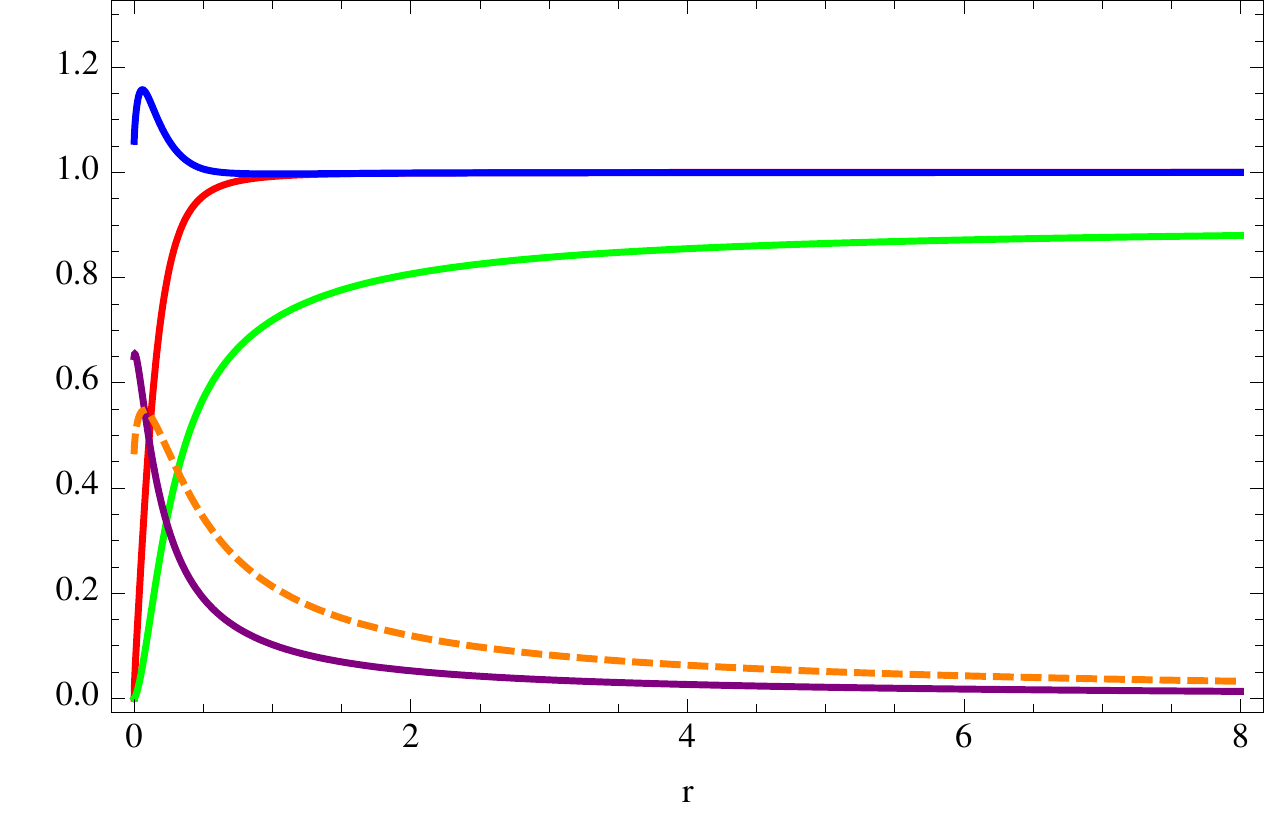}
\end{tabular}
\caption{\small An example of the background of the single-edge hairy electron star solution. {\it Left:}  fluid parameters as functions of the radial coordinate $r$: $\hat{\rho}$ (red), $\hat{n}$ (green), $\hat{p}$ (blue). It is easy to see that these functions are not monotonic along the radial coordinate as in the pure electron star case. {\it Right:} metric and scalar fields of the hairy electron star as functions of the radial coordinate $r$: $f/c^2r^2 $(red), $g r^2$(blue), $h/c\mu$ (green), $\mu_{\text{loc}}$(orange), $\hat{\phi}$ (purple).}
\label{hes}
\end{center}
\end{figure}

Outside the star 
the system is described by Einstein-Maxwell-Scalar gravity alone
(\ref{eomhs1} - \ref{eomhs4}), as in the pure holographic superconductor.
At the boundary of the star $r_s$, we need to match the solution to
the full fermion-plus-boson system to the pure boson system. This
implies the boundary conditions for the outside region: 
\bea && f(r_{s+})=f(r_{s-}),~~ g(r_{s+})=g(r_{s-}),~~ h(r_{s+})=h(r_{s-}),\nonumber\\
&&h'(r_{s+})=h'(r_{s-}),~~ \hat{\phi}(r_{s+})=\hat{\phi}(r_{s-}),~~
\hat{\phi}'(r_{s+})=\hat{\phi}'(r_{s-}).
\eea 
We then integrate from $r_s$ to the boundary using the equations of
motion in (\ref{eomhs1} - \ref{eomhs4}). 
For the values quoted above: we find after integration the $AdS_4$
boundary values $(c, \Phi_1,\mu,Q, E)\simeq (1.104, 0.106, 0.267, 0.062, 0.011)$,  and therefore $F=E-\mu Q\simeq-0.005.$ For the grand canonical ensemble 
it follows that $F/\mu^{3}\simeq-0.289$.
Fig. \ref{hes} shows the way that the hairy electron star solutions behave.
$f,~g,~h,~\phi$ and the fluid parameters of fermions for the parameter
above. 

{\bf CASE II: hairy ES with two edges:}
departing from the hairy electron star with a single edge and a fermi fluid core
in the interior, a different star evolves upon increasing $\hat{m}_f$ while
the other parameters are kept fixed. The reason is that the local chemical
potential in the holographic superconductor background increases first
as one moves outward, before it starts to decrease. This means that
when $\hat{m}_f$ becomes bigger than the local chemical potential in
the deep interior at the horizon, it is no longer possible to support
a fermi-fluid near the horizon. 
Instead
an inner edge will arise where fermions  start to materialize. As a typical example for this case, consider  
$(\hat{m}_b^2, \hat{q}, \hat{u}, \hat{m}_f, \beta)=(-2, 1.55,  6, 0.725, 19.951)$. 
Since there is no fermi fluid possible in the interior, the near horizon
geometry has to be the same as the holographic superconductor. 
The inner edge of the star $r_{s1}$ is defined as (\ref{staredge});
for the quoted values this is at $r_{s1}\simeq 0.064.$ Then at
$r_{s1}$ we connect to the interior of the star where the system is
described by full combined fermi-boson system eqn. (\ref{hes1} - \ref{hes4}) until it runs to the second edge of the star. 
Beyond this edge 
the system is again fluid-less and described by Einstein-Maxwell-scalar 
gravity (\ref{eomhs1} - \ref{eomhs4}). Here the outer edge is at
$r=r_{s2}\simeq 0.879$. At the asymptotical $AdS_4$ boundary, we
obtain the values $(c, \Phi_1,\mu,Q, E)\simeq (3.856, 1.237, 3.026, 7.667, 15.496)$,  thus 
$F/\mu^{3}\simeq-0.279$. 
In Fig. \ref{hes2} we show
the functions $f,~g,~h,~\hat{\phi}$ for the fluid parameters 
of the specific example mentioned in the above.
\begin{figure}[h] 
\begin{center}
\begin{tabular}{ccc}
\includegraphics[width=0.45\textwidth]{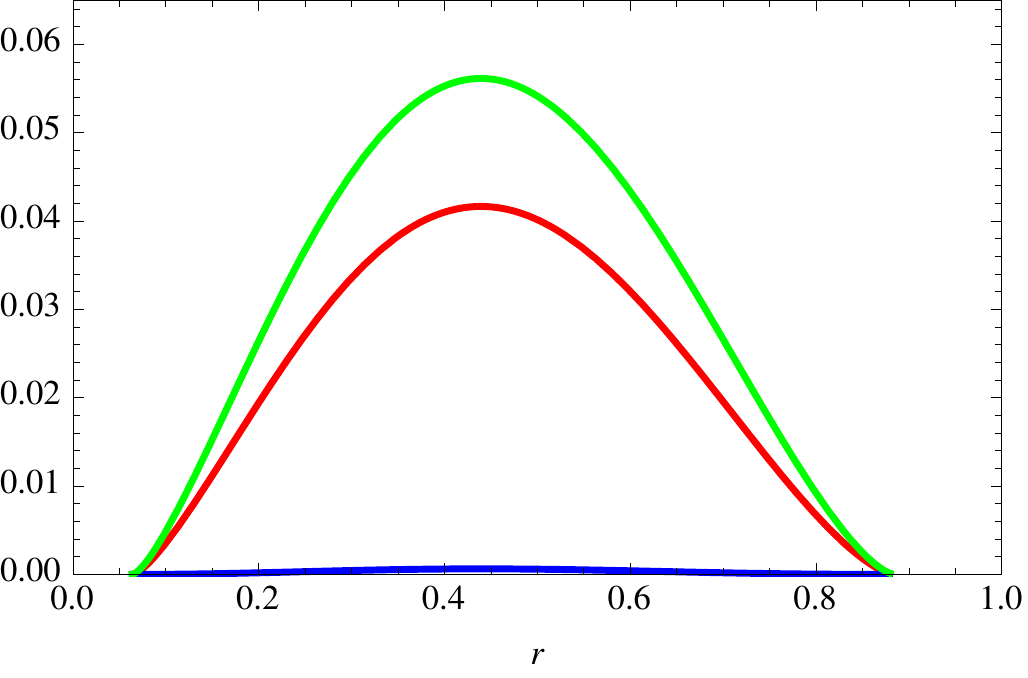}
\includegraphics[width=0.45\textwidth]{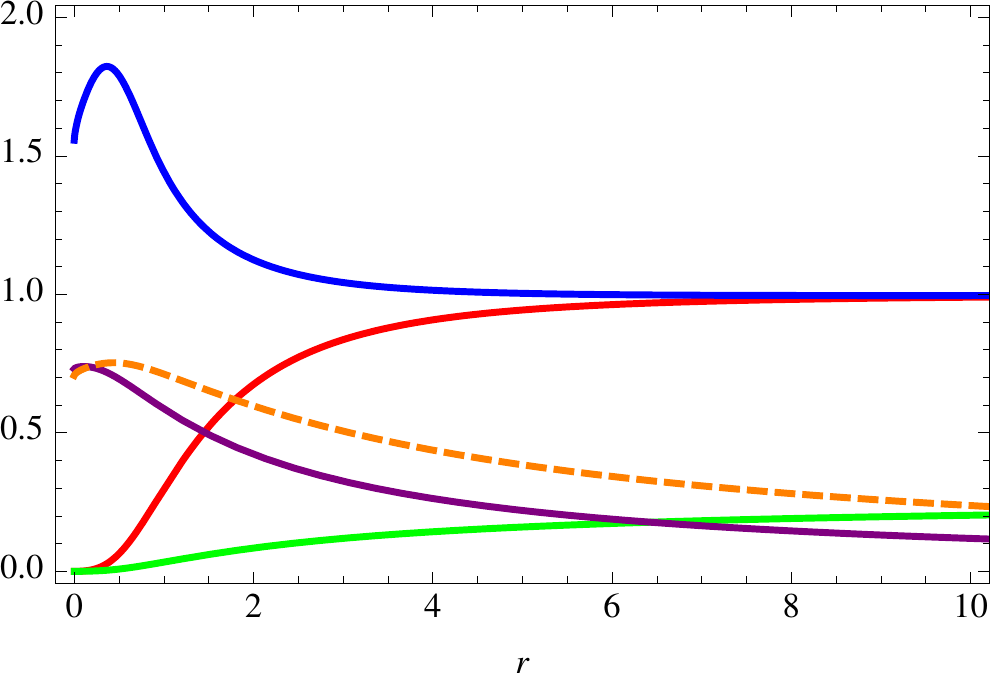}
\end{tabular}
\caption{\small An example of the background of the double-edge hairy electron star solution. {\it Left:} fluid parameters as functions of the radial coordinate $r$: $\hat{\rho}$ (red), $\hat{n}$ (green), $\hat{p}$ (blue). We can see that there are two edges. {\it Right:} metric and scalar fields of the hairy electron star as functions of the radial coordinate $r$: $f/c^2r^2 $(red), $g r^2$(blue), $h/c\mu$ (green), $\mu_{\text{loc}}$(orange), $\hat{\phi}$(purple).}
\label{hes2}
\end{center}
\end{figure}

Regarding the hairy electron star as a fermi fluid living on a
holographic superconductor background, it is also clear that when the fermion mass $\hat{m}_f$ becomes larger than the maximum value of the local chemical potential of the holographic superconductor background, there will be no hairy star solution anymore.

\bigskip

{\bf Holographic Luttinger's theorem:}
In principle the non-zero order parameter corresponding to non-zero
vev of the scalar field signals that the field theory state dual to
the hairy electron star is in a symmetry broken state. Therefore the standard
Luttinger theorem need not apply. Of course since we are essentially
describing a system of non-interacting bosons and fermions, there is a 
simple variant of the Luttinger relation. 
In holography one finds that the total field theory charge density
equals
\be
Q_{\text{FT}}= Q_{\text{bulk}}+Q_{\text{horizon}}
\ee
where $Q_{\text{bulk}} = \sum_{\text{occupied~charged~states}} Q^{(i)}$.
For the pure electron star solution, where the only bulk constituents
are the charged fermions the field theory Luttinger's theorem, which
states that the volume of the Fermi surface is equal to the charge
density of the fermions, follows from the bulk Luttinger's theorem. In
Lifshitz spacetimes 
there is no contribution of horizon charge
\cite{{Hartnoll:2011dm},Iqbal:2011bf} and thus the boundary charge
density is equal to the bulk charge density. 
Due to the fluid approximation there will be a set of infinitely many
Fermi surfaces (one for each radial mode) whose fermi momentum is the same in the bulk and in the
dual field theory \cite{Hartnoll:2011dm, {Iqbal:2011in},
  {Cubrovic:2011xm}}. Therefore the Luttinger relation in the bulk and
the boundary is the same.
For the hairy electron star, the near horizon geometry is still
Lifshitz with a finite $z$, so there is again no contribution of
horizon charge. Also due to the fluid approximation there is a similar
set of infinitely many
Fermi surfaces corresponding to occupied radial modes.
However, the HES solution corresponds to a condensed state with vacuum
charge $Q_b$. But this effect can easily be accounted for. It is
straightforward to see that we should now have a Luttinger relation \be \sum_n\frac{2}{(2\pi)^2}V_n=Q-Q_b,\ee 
where $V_n=\pi k_F^{(n)2}$ is the volume of the $n$-th Fermi surface
of the dual field theory and $Q_b$ the boson charge density
$Q_b=\int_0^\infty dr\sqrt{-g}2\hat{q}_b^2\hat{\phi}^2h/f$. This is
familiar from fractionalized Fermi systems \cite{Huijse:2011hp} 
in the broken phase. As a reservoir of charge the Bose condensate now takes
over the status of the horizon charge.

\section{$T=0$ Phase diagram}
\label{phasediag}

With the exact solutions in hand we can now construct the phase
diagram at zero temperature in detail. As mentioned above, we will
keep fixed $\hat{u}=6$, $\hat{\beta}\simeq19.951$ and we rescaled $q_f$
to be 1. The three tunable parameters we consider are $\hat{m}_f$,
$\hat{m}_b$ and $\hat{q}_b$. 

\subsection{Quantum phase transition boundaries}
Assuming that all the phase transitions are continuous lines of
instability we determined in Fig. \ref{figBF-bounds} should correspond to the
exact phase boundaries. This was already known for the AdS-RN/ES phase
boundary for $\hat{m}_b^2 \gg 1$; in the fluid limit the nature of the
transition is not known yet, but it is assumed to be
continuous.\footnote{At finite $T$ it is 3rd order
  \cite{Puletti:2010de,Hartnoll:2010ik}.} Also the continuous
  AdS-RN/HS phase boundary equals the instability curve.  This has
  been studied explicitly in \cite{Iqbal:2010eh,{Faulkner:2010gj},{Iqbal:2011aj}} 
for $\hat{q}_b=0$ and it is BKT type. For finite $\hat{q}_b$ it should still be
BKT and the only difference is that the near horizon geometry for the
condensed phase is $\widetilde {\text{AdS}}_2$\footnote{It is an AdS$_2$ region with a vev for the scalar field. $\widetilde {\text{AdS}}_2$ is used to distinguish the near horizon AdS$_2$ geometry of RN without scalar vev. } at $\hat{q}_b=0$ while
Lifshitz at finite $\hat{q}_b$. For the new phase boundaries adjacent to the
new Hairy Electron Star phase, 
there was however a puzzle that the
fermi-instability line in the holographic superconductor phase did not
smoothly join the Fermi-instability line in the AdS-RN phase.

\begin{figure}[h]
\begin{center}
\begin{tabular}{ccc}
\includegraphics[width=0.7\textwidth]{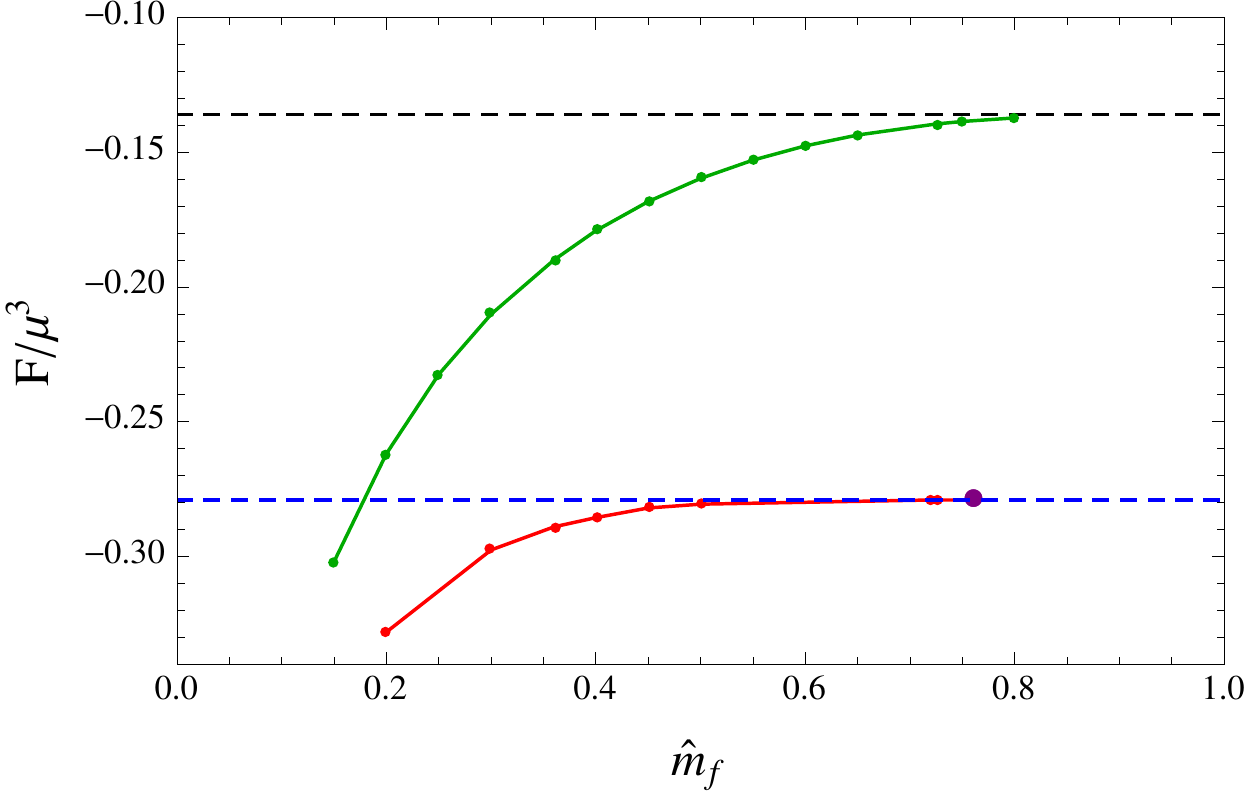}
\end{tabular}
\caption{\small {\it Free energy of the four solutions and the phase
    transition between HES and HS for fixed scalar mass and charge
    with changing fermion mass}. Free energy $F/\mu^3$ vs. $\hat{m}_f$ with
  $(\hat{m}_b^2, \hat{q}_b, \hat{u}, \beta)=(-2, 1.55, 6, 19.951)$:  RN
  (black), ES (green), HS (blue), HES (red). The free energy of RN or
  HS does not depend on $\hat{m}_f$. The purple marked point is the
  bound where we can find hairy ES solution. Here we choose
  alternative quantization for the scalar field in HS and HES
  cases. 
For standard quantization the plot is qualitatively the same because this picture only concerns the transition between HES and HS while not 
the transition between ES and HES. In other words, 
the instability shown in this plot is only related to fermions in the HS background, so there is no qualitative dependance on the boundary condition for the scalar field.
 There is also no qualitative difference for other boundary conditions of the scalar field.}
\label{phase1}
\end{center}
\end{figure}

We tested the continuity of the phase transition of these new phase
boundaries between HS/HES and HES/ES as well as the exact
location of the phase boundary by computing the free energies of the
exact solutions along the section $\hat{m}_b^2=-2$. (Other values of $\hat{m}_b^2$
are numerically hard to control).
The resulting free energy of the four kinds of solutions for
$\hat{q}_b=1.55$ as a function
of $\hat{m}_f$ is given in Fig. \ref{phase1}. 
%
What this study of the free-energy reveals is that the phase boundary
between the HS and the HES solutions is {\em not} given by the
deformed BF bound for fermions in the HS background. With the
knowledge how the HES 
is constructed it is easy to see why. As we
lower $\hat{m}_f$ from the HS reason, there is a critical value
$\hat{m}_f \sim \mu_{\text{loc}}^{\text{max}}$ where the fluid first
forms, {\em but due to the non-monotonic behavior of
  $\mu_{\text{loc}}$ it does not form at the Lifshitz horizon but in
  the interior.} The first star one encounters by tuning $\hat{m}_f$
down is the two-edged HES. The BF instability bound for fermions,
however, is constructed from the Lifshitz scaling solution. Indeed
tracing the sequence of solutions, one finds that at this value of
$m_f$ one has a crossover from the two-edge HES to the single-edge
HES.

For the free energy at value $\hat{m}_b^2=-2, \hat{q}_b=1.55$ the HES remains the preferred phase all the way down to $\hat{m}_f=0$. However, for a higher $\hat{m}_b^2$ or lower $\hat{q}_b$ 
one will encounter
the transition line between the single edged HES and the ES
solutions as one keeps lowering $\hat{m}_f$. We shall explain below that this exactly corresponds with the
 BF bound of bosons in the ES near horizon region
 $\hat{m_b}^2-h_0^2\hat{q_b}^2=-\frac{(z+2)^2}{4g_0}$ (for the standard quantization of the scalar boundary
 condition.)

\begin{figure}[h]
\begin{center}
\begin{tabular}{cc}
\includegraphics[width=0.7\textwidth]{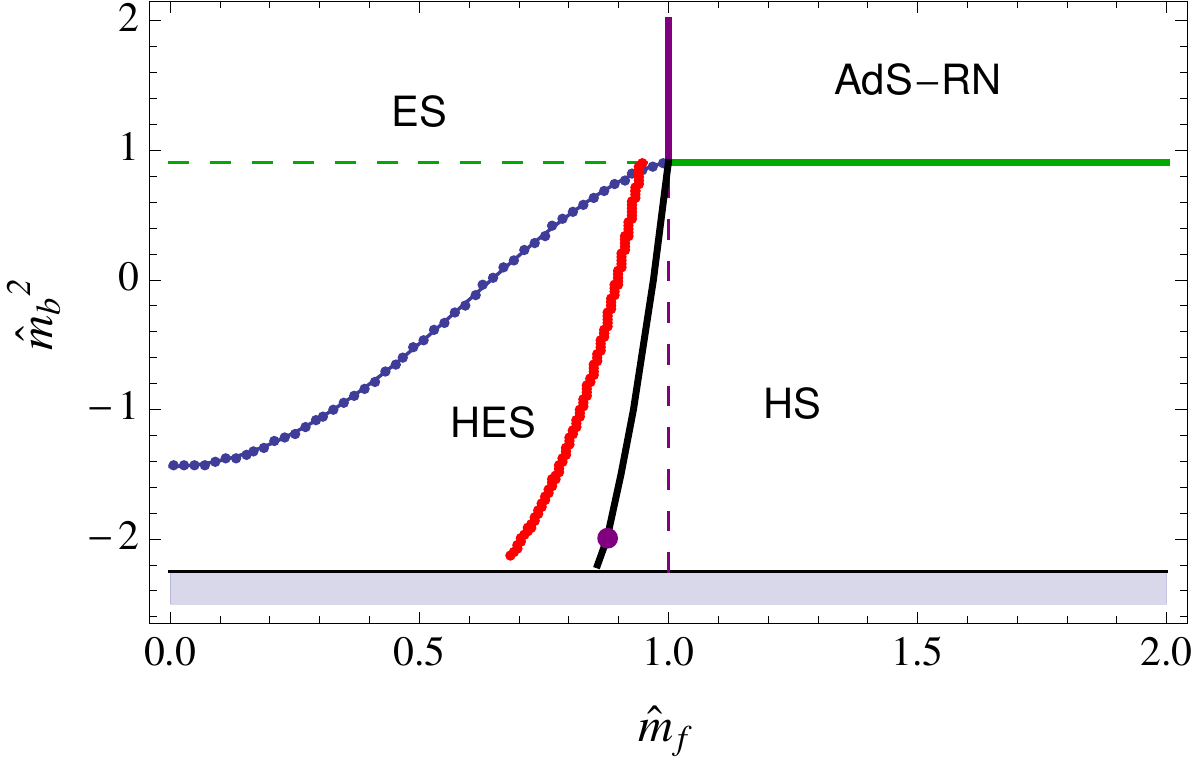}
\end{tabular}
\caption{\small The improved phase diagram of Fig. \ref{figBF-bounds} for $(\hat{q}_b,\hat{u},\beta)\simeq (1.55,6,19.951)$.  
Here the red line is the incorrect instability curve from analysis of  fermions in the  near horizon of HS. The black line denotes the actual instabilities of fermions in HS. 
Due to numerical difficulties, this black line is only a rough illustration. The purple marked point is the transition point between HES and HS in the standard quantization case which is qualitatively the same to Fig. \ref{phase1}.}
\label{figBF}
\end{center}
\end{figure}

In Fig. \ref{figBF} we show these results. The red dotted line in
Fig. \ref{figBF} is the fermion instability bound, which we now know
denotes the transition between the one-edge HES  and the two-edge
HES. 
The real
quantum phase transition between a HS and two-edge HES happens at the
black line in the left figure. This black line can be obtained
numerically by demanding $\hat{m}_f=\max[\mu_{\text{local}}(r)]$. It
will also depend on the boundary condition of the scalar field but the
dependence is only quantitative which is indicated from Fig. \ref{bghs}.

With this understanding of the phases for fixed $\hat{q}_b$, we can
change this parameter as well. Nothing changes
qualitatively. Specifically as one changes $\hat{q}_b$ one finds
that

\begin{itemize}
\item[1,] At $\hat{m}_f=0$, the value of $\hat{m}_b^2$ on the ES-HES
  phase boundary line is always smaller than the value of
  $\hat{m}_b^2$ for the HS-RN boundary as the BF bound in the ES
  background is always lower than the BF bound in the AdS$_2$
  background. This means that the ES-HES boundary always exists and obeys the relation (\ref{estohesbfbound}) no matter how $\hat{q}_b$ changes.
\item[2,] One important characteristic is that for $\hat{m_b}^2$ near to or
  equal to the unitarity bound (the bottom of the phase diagram), the
  single-double edge HES crossover is always on the right of the
  ES-HES transition. This can be seen directly from a 3D plot of the
  BF stability diagram where $\hat{q}_b$ is taken into account. It can
  also be seen from the fact that the single-double edge crossover can
  only reach $\hat{m}_f=0$ for $\hat{q}_b>2$ while the ES-HES line
  always intercepts $\hat{m}_f=0$. From (\ref{estohesbfbound}) and
  (\ref{nhlif}) we see that this intercept value of $\hat{m}_b^2$ for
  the ES-HES transition at $\hat{m}_f=0$ is always larger than that of
  single-double crossover. Furthermore it is immediately obvious from
  the construction that the double edge HES-HS transition is always to
  the right of single-double HES crossover because the double-edge HES
  solution exists earlier than the single-edge solutions when we
  decrease $\hat{m}_f$ from the HS side. These two facts together
  imply that the ES-HES transition line is always above the HES-HS transition for all the possible parameter regions and the HES region always exists. There is no direct phase transition from ES to HS except at the critical point.
\end{itemize}




\bigskip
By construction we have quite a good understanding of the transition
from the holographic superconductor to the hairy electron star.
Let us finally discuss the other transition between the electron star and
the hairy electron star in more detail.
From the phase diagram Fig. \ref{figBF} we can clearly see that the transition
between ES and HES can happen  both directly to the one-edge HES
solution as well as to the two-edge HES solution because the ES-HES
transition line intersects with the IR stability bound signaling the
single-double edge crossover. These two ES-HES transitions are the same in nature but a little different in detail.
\vspace{.3cm}\\
{\bf CASE I: Phase transition between ES and one-edge HES:}
In all cases with a non-trivial bosonic background the instability to forming scalar hair is a 
perturbative instability: the scalar perturbations on the geometry
background become tachyonic modes and render the background unstable
towards a new ground state. At finite temperature, even when backreactions of the scalar field to the geometry are considered, at the onset of the phase transition the scalar field is very small in the spacetime and can be treated as perturbations, so the perturbative analysis is still valid and the transition is very continuous (second order). 

At zero temperature, on the other hand, one encounters the following
puzzle. After taking backreaction into account, the near horizon value
$\phi_0$ is always non-zero 
--- it is located at the symmetry breaking minimum of the quartic
potential. However, just before the transition point under any
boundary condition for the scalar field, 
$\phi$ 
should vanish for a continuous phase transition, and this seems to
contradict with the fact that $\phi_0$ just after the transition is
distinctly non-zero at the horizon. 

As explained in \cite{{Iqbal:2010eh},Iqbal:2011aj}, the resolution is
that there exists a special emergent IR scale for the system slightly below the BF bound
and the condensate only has a finite effect below this IR scale. Near
the phase transition, the dual field theory order parameter extracted
from the AdS boundary value of the condensate is infinitesimal. 
In fact the value stays very small for a long distance along the radial
direction into the interior until it reaches this special IR scale and
the bulk condensate starts to become of finite size.

For the standard holographic superconductor this IR scale
\cite{{Iqbal:2010eh},Iqbal:2011aj} is 
\be 
\Lambda_{\text{IR}} \sim
\mu\exp{\bigg(-\frac{c_0}{\sqrt{\Delta_c-\Delta}}\bigg)}, ~~~~~c_0=\pi
\sqrt{\frac{d(d-1)}{2\Delta_c-d}},
\ee 
where $\Delta$ is the UV scaling dimension
and $\Delta_c$ is the critical UV scaling dimension at the BF bound.
This scale is exponentially small close to the BF bound, so the finite
$\phi$ effect is only constrained in an exponentially small distance 
in the radial direction (one can think of $\Lambda_{IR}$ as the distance to the center of AdS).  
  This makes the difference in
the free energy also exponentially small, so this resolves the
contradiction between a finite jump in the horizon value and a
continuous phase transition between the HS and the AdS-RN black hole. 
In fact it shows that 
the phase transition is of the most continuous kind: it is a BKT phase
transition. It allows one to think of the AdS geometry in the
following way.  In the exponentially near horizon region below this IR
scale the geometry is the backreacted Lifshitz geometry, but it soon
turns into an almost unaffected near horizon AdS-RN AdS$_2$ geometry 
above $\Lambda_{{\text IR}}$, so there exists an intermediate
``semi-local quantum critical'' region with the geometry of AdS$_2$ in an intermediate scale $\Lambda_{\text{IR}}$ to $\mu$.


\begin{figure}[h]
\begin{center}
\begin{tabular}{ccc}
\includegraphics[width=0.45\textwidth]{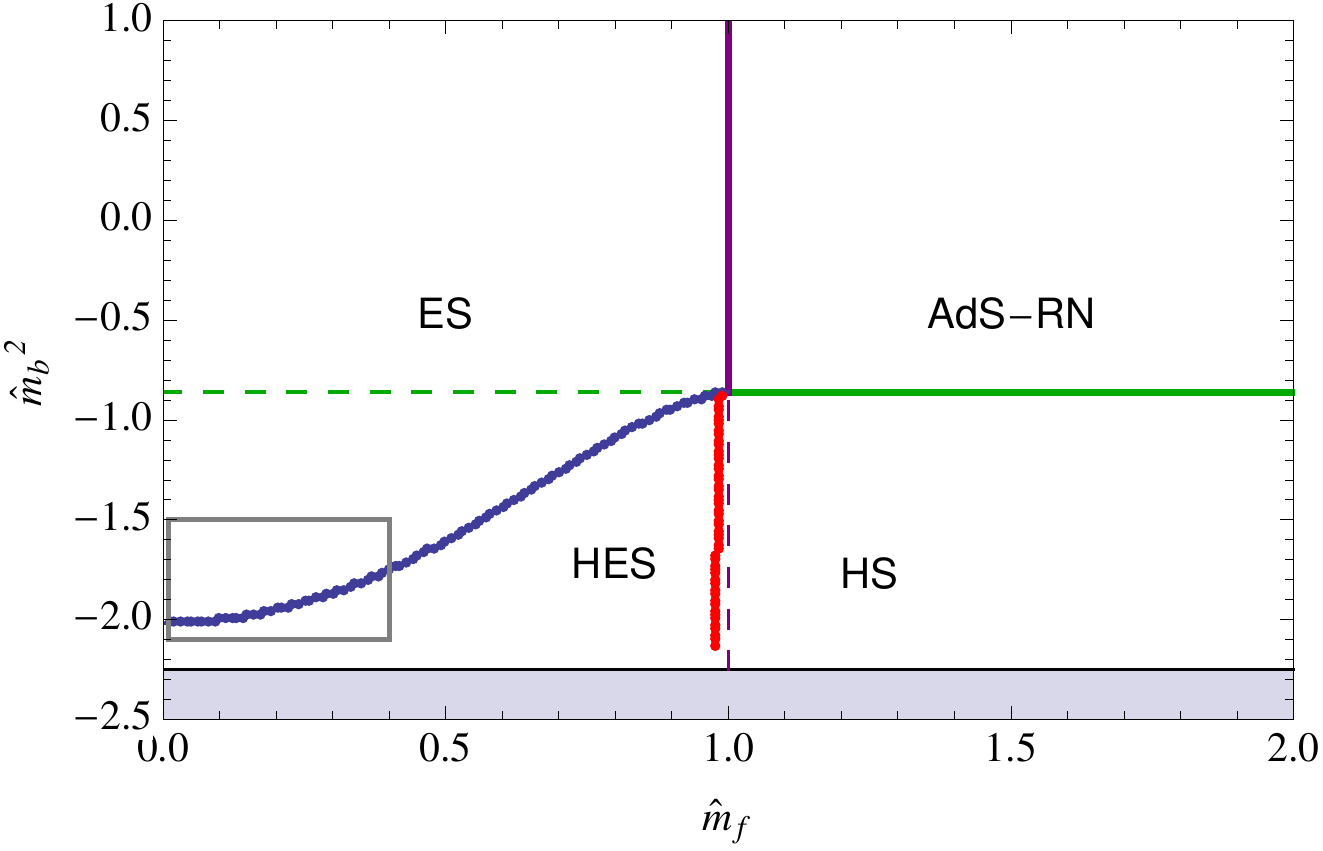}
\includegraphics[width=0.45\textwidth
]{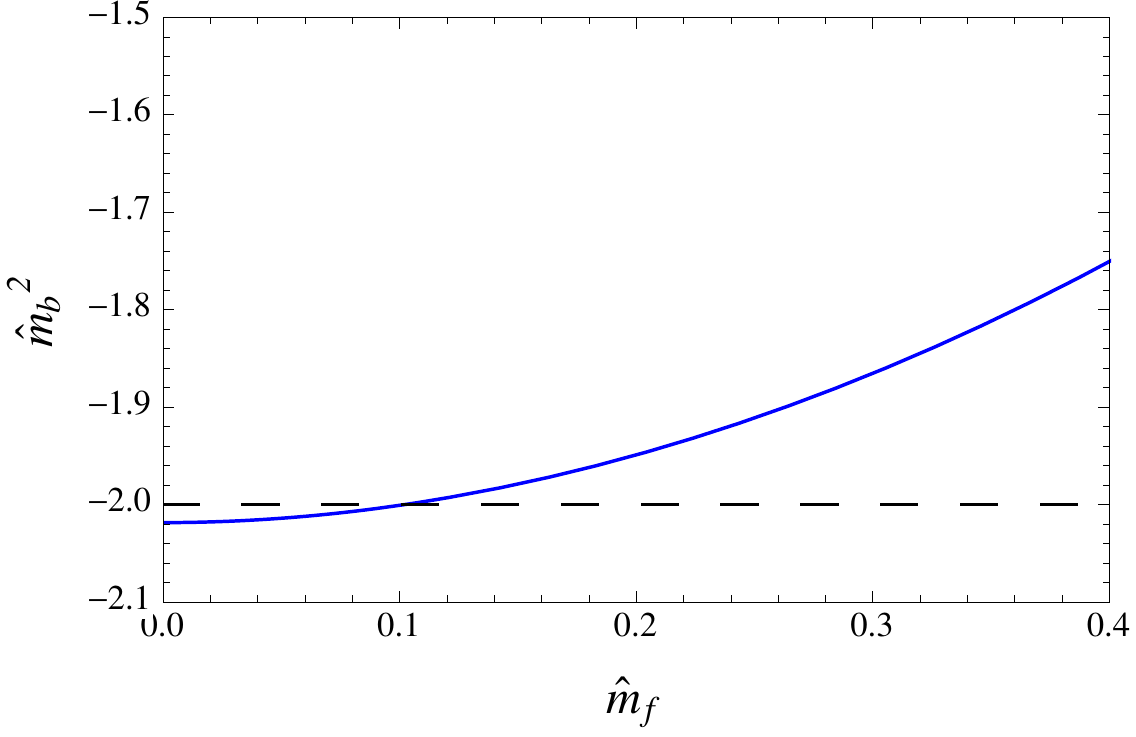}
\end{tabular}
\caption{\small {\it Left:} phase diagram in the
  $\hat{m}_b^2$-$\hat{m}_f$ plane for $\hat{q}_b=0.8$; {\it Right:} a clearer picture for the blue curve of the gray box region in the left figure. This is qualitatively the same as Fig. \ref{figBF}. The red line is the one-edge HES boundary.}
\label{phasediagramq8}
\end{center}
\end{figure}

We now show that the phase transition from HES to ES has the same
feature as the transition from HS to AdS-RN. Numerically we show that
near the BF bound with the standard boundary condition for the scalar
field there indeed exists a intermediate region just beyond the
near-horizon Lifshitz region where the scalar field drops to almost
zero and the geometry in that intermediate region becomes the near
horizon geometry of ES. This confirms that there exists an emergent IR
scale below which the effect of the condensation becomes significant. 

\begin{figure}[h]
\begin{center}
\begin{tabular}{cc}
\includegraphics[width=0.45\textwidth]{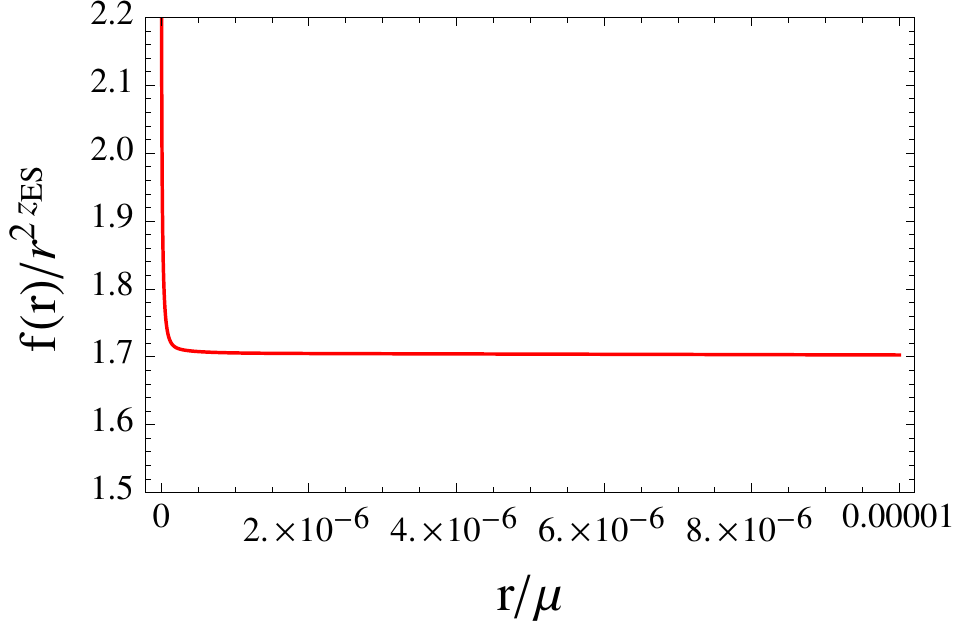}
\includegraphics[width=0.45\textwidth]{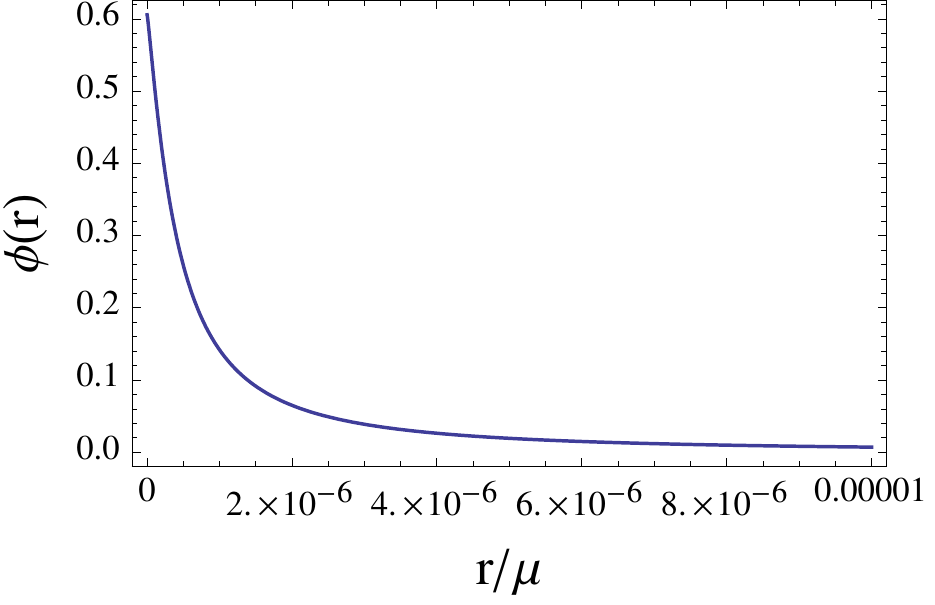}
\end{tabular}
\caption{\small Evidence of the existence of an intermediate critical region for $\hat{m}_b^2=-2$, $\hat{q}_b=0.8$ and $\hat{m}_f$=0.12. The function of $f$ in the metric behaves as $r^{2z_{{\text ES}}}$ in an intermediate region of $r/\mu$. }
\label{criticalreg}
\end{center}
\end{figure}

To avoid numerical complexities we choose $\hat{q}_b=0.8$; the phase
diagram for this case the phase diagram is given in Fig. \ref{phasediagramq8}. In Fig. \ref{criticalreg} we 
show the existence of the critical region for
$(\hat{m}_b^2,\hat{q}_b,\hat{m}_f)=(-2, 0.8, 0.12)$ and
$(f_1,f_2)=(9.123, -10)$. We can see when moving
outward one extremely rapidly enters an intermediate region where the
blackening function $f(r)$ behaves as $r^{2z_{\text{ES}}}$ where
$z_{\text{ES}}$ is the Lifshitz scaling exponent of the electron star
solution at this $\hat{m}_f$. The IR scale $r/\mu$ at which the
critical regions starts to exist should be related to  how close it is to the transition point and
to the value of the condensate $\langle \mathcal{O}_2\rangle$ but here
due to numerical reasons we choose the initial value of $f_1$ and
$f_2$ a little away from the transition point so that the critical
region starts to exist at a scale $r/\mu$ around $10^{-6}$. 

In our case, the IR critical scale \cite{Iqbal:2010eh, {Edalati:2011yv}}  is \be \Lambda_{\text{IR}}\sim \mu
\exp{\bigg(-\frac{\pi}{\sqrt{g_0(m_c^2-\hat{m}_b^2)}}\bigg)},\ee where
$m_c^2=\hat{q}_b^2h_0^2-\frac{(z+2)^2}{4g_0}$ is the mass square of
the BF bound (\ref{estohesbfbound}).
It is easy to check that the difference in free energy compared
to the pure electron star background should therefore scale as 
\be \label{fediff}\delta F\sim \mu^3
\exp{\bigg(-\frac{(z+2)\pi}{\sqrt{g_0(m_c^2-\hat{m}_b^2)}}\bigg)}.\ee

\begin{figure}[h]
\begin{center}
\begin{tabular}{cc}
\includegraphics[width=0.5\textwidth]{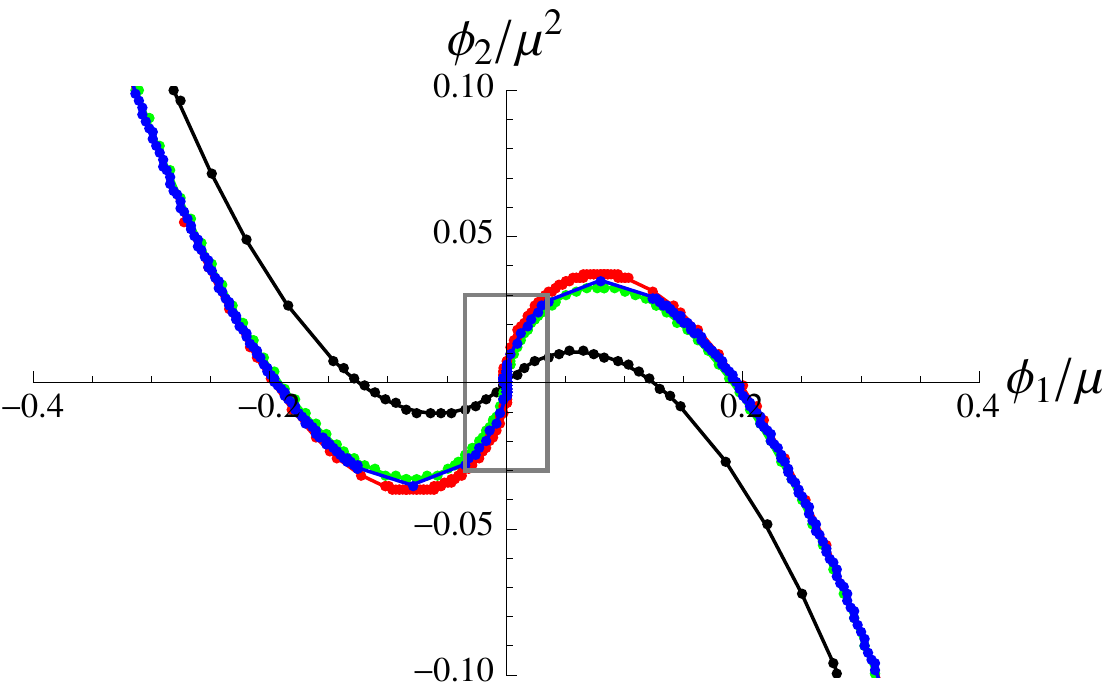}
\includegraphics[width=0.5\textwidth]{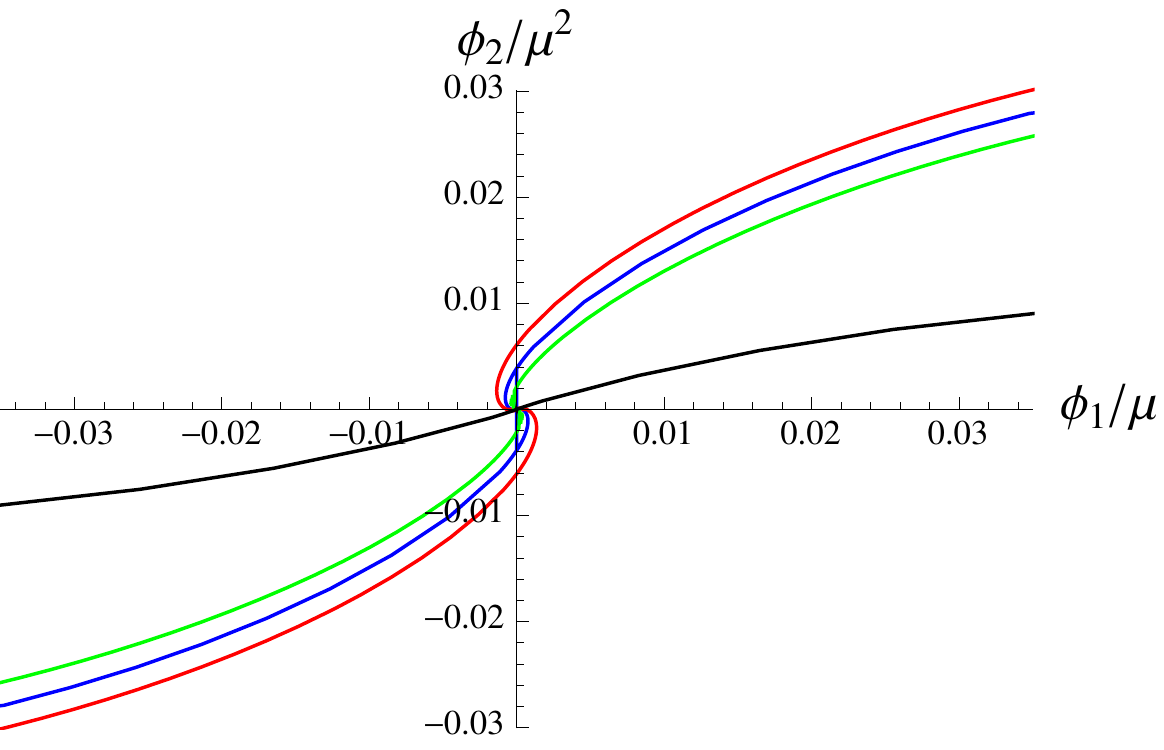}
\end{tabular}
\caption{\small Plots of $\Phi_2/\mu^2$ as a function of $\Phi_1/\mu$ in the condensed phase (hairy ES background). Different points on each curve correspond to different values of the shooting parameters. THe right plot is the zoomed in version of the left gray  boxed region. Here 
$\hat{m}_f=0.6$(red), $0.55$(blue), $0.5$(green), $0.05$(black),  correspondingly, $g_0(m_c^2-\hat{m}_b^2)/(z+2)^2=0.074,0.061,0.046, -0.002$.}
\label{figalphabeta}
\end{center}
\end{figure}

We can check this explicitly. 
We focus on the line of $\hat{m}_b^2=-2$ in Fig. \ref{phasediagramq8} and vary $\hat{m}_f$ to change the BF bound $m_c^2$. In Fig.\ref{figalphabeta}, we plot out the curve generated by boundary values of $\Phi_1$ and $\Phi_2$ when we change the boundary values of $f_1$ and $f_2$. The different curves correspond to different values of $\hat{m}_f$ for fixed $\hat{m}_b^2=-2$ and $\hat{q}_b=0.8$, which correspond to 
$g_0(m_c^2-\hat{m}_b^2)$ ranging from $-0.019$ to $2.368$. For
standard quantization, normalizable solutions correspond to the
intersecting point of the curve with the $\Phi_2$ axis while for
alternative quantization, normalizable solutions correspond to the
intersecting point of the curve with the $\Phi_1$
axis. 
From the left figure we can see that the curves always intersect with
the $\Phi_1$ axis far from the origin, but our interest is the
intersection with the $\Phi_2$ axis near the origin. It is numerically
difficult to get very close to the transition point, but zooming in
--- displayed in the right figure --- we directly see that when
$m_c^2-\hat{m}_b^2$ decreases the expectation value of $\Phi_2$ will
also decrease. It is expected that $\Phi_2$ goes to zero when $\hat{m}_b^2$ approaches exactly $m_c^2$.

\begin{figure}[h]
\begin{center}
\begin{tabular}{cc}
\includegraphics[width=0.45\textwidth]{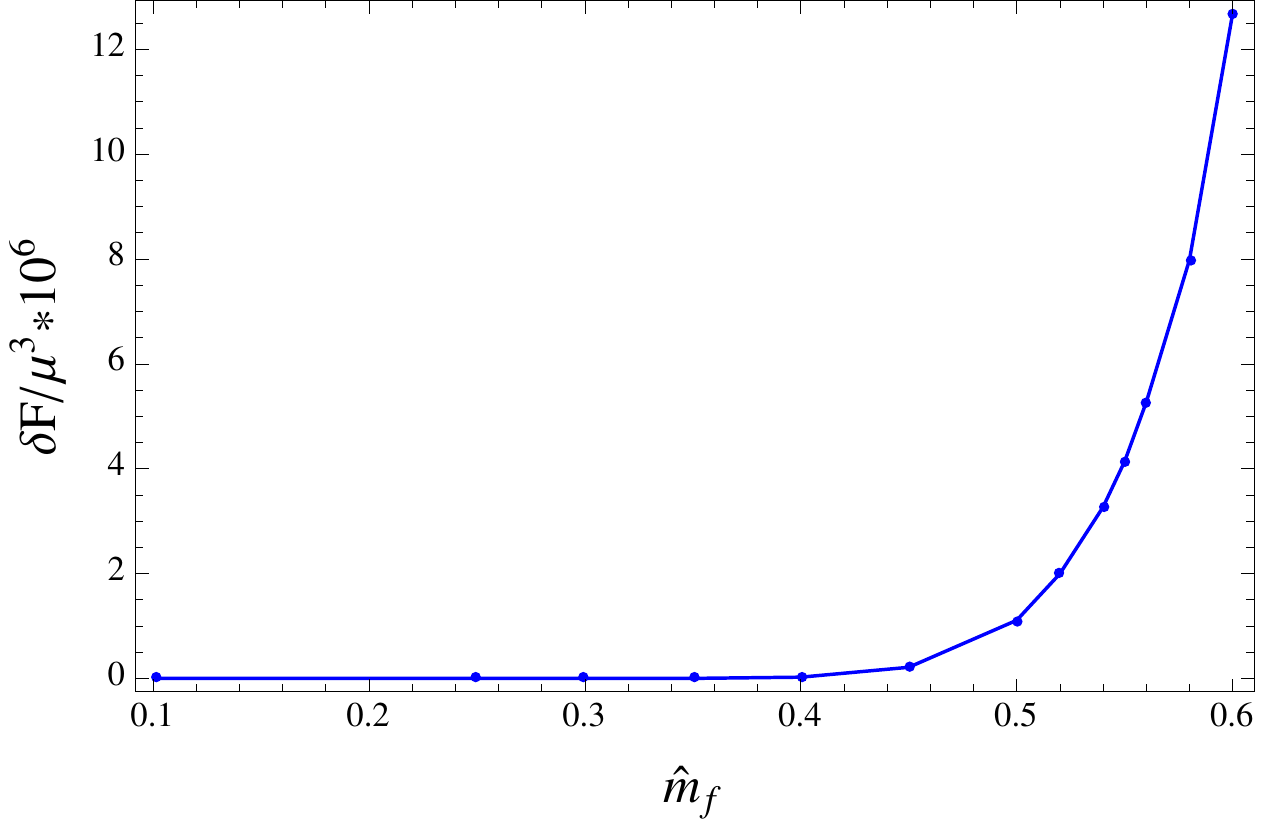}
\includegraphics[width=0.4\textwidth]{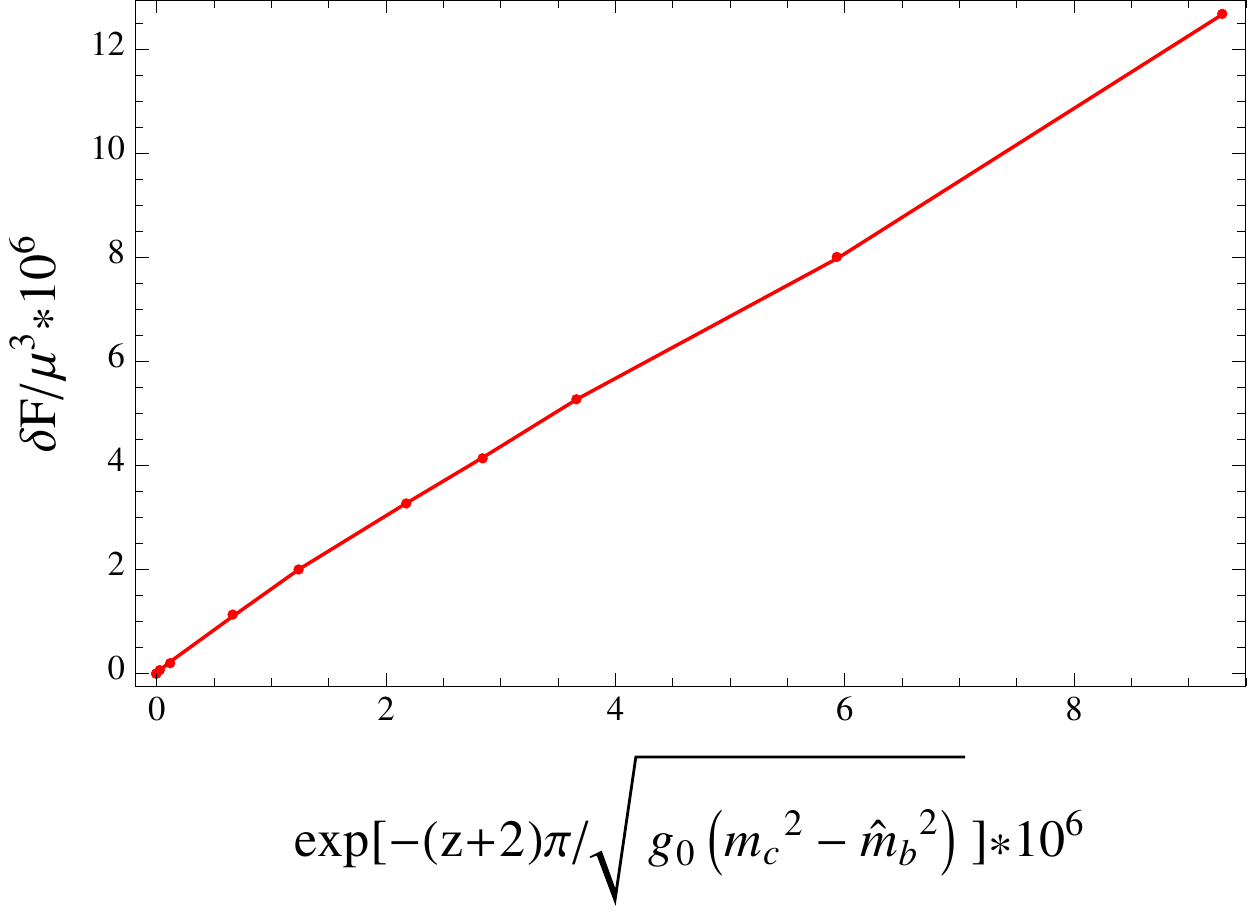}
\end{tabular}
\caption{\small Difference of free energy $\delta F$ as a function of
  $\hat{m}_f$ (left) and of
  $\exp{(-\frac{(z+2)\pi}{\sqrt{g_0(m_c^2-\hat{m}_b^2)}}})$ (right)
  for $(\hat{m}_b^2, \hat{q}_b, \hat{u}, \beta)=(-2, 0.8, 6, 19.951)$
  and standard boundary condition for the scalar field.}
\label{figdeltaf}
\end{center}
\end{figure}

\begin{figure}[h]
\begin{center}
\begin{tabular}{ccc}
\includegraphics[width=0.7\textwidth]{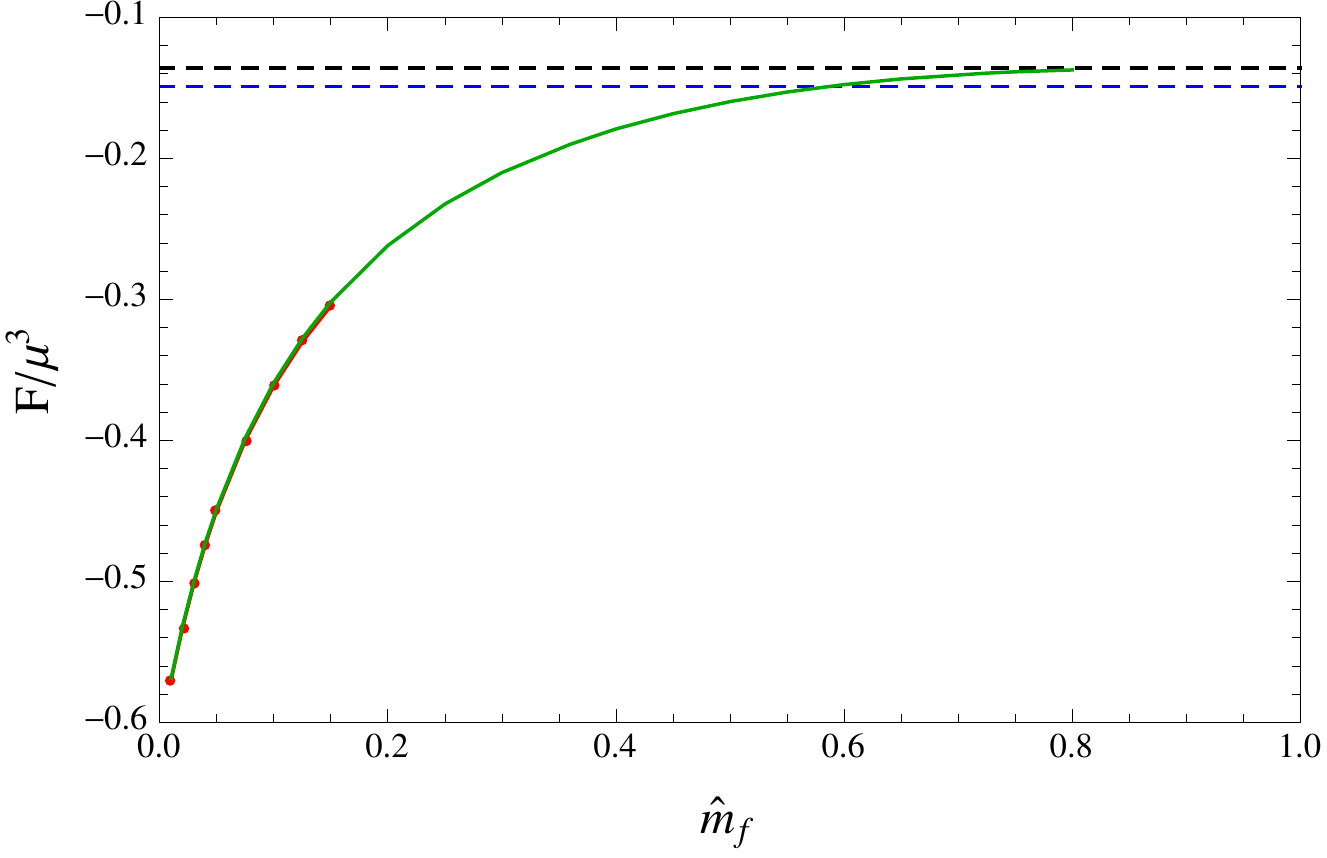}
\end{tabular}
\caption{\small {Free energy of HES and ES solutions for alternative boundary condition.} The plot $F/\mu^3-\hat{m}_f$ with $(\hat{m}_b^2, \hat{q}_b, \hat{u}, \beta)=(-2, 0.8, 6, 19.951)$,  RN (black), ES (green), HS (blue), HES (red): the free energy of HES always lower than ES.}
\label{phase2}
\end{center}
\end{figure}

The difference of the free energy of the two phases as a function of
$\hat{m}_f$ are shown in Fig. \ref{figdeltaf}. We can see from the right
figure in Fig. \ref{figdeltaf} that the free energy difference scales
exactly as (\ref{fediff}). Thus this phase transition is still BKT.

Note that the boundary condition for the scalar field is crucial to
the behavior of the phase diagram. We can see from
Fig. \ref{figalphabeta} that near the BF bound we can still find
solutions with alternative boundary conditions for the scalar field
with a finite expectation value $\Phi_1$. 
In fact
this should be the preferred solution --- 
in general the largest vev of either $\Phi_1$ or $\Phi_2$ is the one with lowest free energy 
\cite{Franco:2009yz}. Already for the pure holographic superconductor, but also for the hairy electron star, the free energy of the solution with alternative boundary conditions is {\em lower} than the free energy of the solution with standard boundary condition. The same conclusion holds in the probe holographic superconductor case \cite{Franco:2009yz}. It is quite puzzling that for a finite density system the alternative one is more stabler as there is at the $1/N$ level a clear double trace deformation where the zero density system should flow to standard quantization. 
We are not sure yet what this implies. Here we simply present the results and defer the answer to future work. 
\vspace{-0.2cm}\\

\begin{figure}[h]
\begin{center}
\begin{tabular}{ccc}
\includegraphics[width=0.8\textwidth]{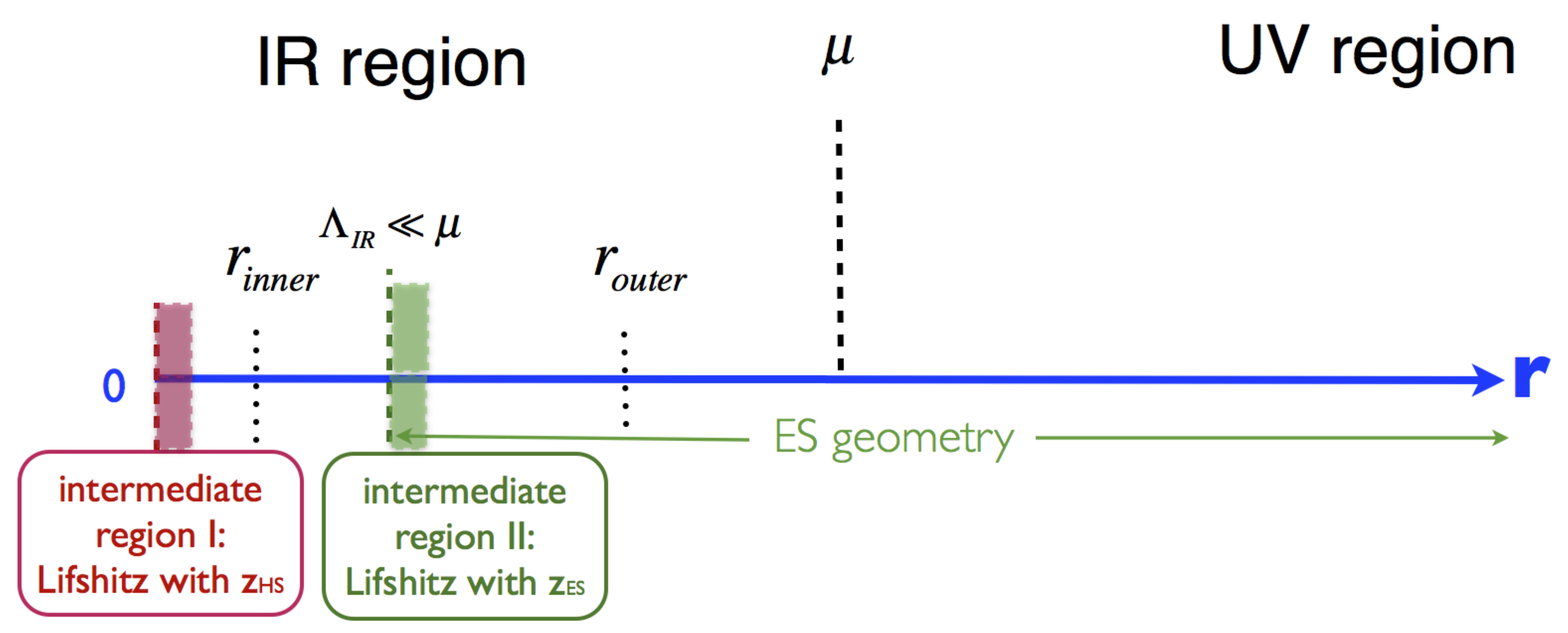}
\end{tabular}
\caption{\small Transition scales for the two-edge hairy electron
  star. The near horizon area of the condensed phase is composed of
  two critical regions and relevant deformations in
  between. Intermediate region I is Lifshitz $z_{\text{HS}}$ and
  intermediate region II is Lifshitz $z_{\text{ES}}$.} 
\label{critical1}
\end{center}
\end{figure}

{\bf CASE II: Phase transition between ES and two-edge HES:}
%
With this knowledge we can be quick in our discussion of 
the phase transition between ES and the two-edge HES. It is  qualitatively
the same in that there are emergent IR scales. In this case moving
inward for the
two-edge HES the condensate starts to be significant near a similar
$\Lambda_{\text{IR}}$ as for the one-edge HES. At this scale the
geometry turns from the near horizon geometry of ES to the geometry of
HES at this scale, which means that the inner edge of the two-edge HES
solutions near the BF bound is also close to 0. At this inner edge,
which can also be thought of as an emergent IR scale the solution
becomes fully HS. This behavior is sketched in Fig. \ref{critical1}.

\subsection{The critical point}

The final noteworthy feature of the phase diagram is the quadruple 
critical point 
at which any two phases can 
be connected with each other. From this critical point, we can in principle realize quantum phase 
transitions directly between AdS-RN and HES or between HS and ES by
traversing diagonally in the phase diagram through the critical point. 
This is quite interesting 
and we will leave this for future study. Here we wish to notice that 
the existence of the emergent IR scales resolves some of the paradoxes
one might encounter when considering this situation. At the critical point, 
the AdS-RN solution is connected to the HS solution, so when 
we decrease $\hat{m}_f$ the phase transition from HS to HES should 
be the same as the phase transition from AdS-RN to ES. But because 
ES is always one-edge, this seems to be in contradiction with the
 fact that the phase transition from HS to HES at the critical point 
is to two-edge HES solution and one edge HES solutions only exist to 
the left of the red curve in Fig. \ref{figBF}. In fact, this ``jump''
is a direct consequence of the discrete difference in 
the 
near horizon geometry between RN and HS at the critical point. We saw
before that this discreteness paradox gets resolved by the exponential
suppression of the condensate away from the horizon. So it goes here.
 Although the ES solution at the critical point is one-edge, the HES solution 
at the critical point is still two-edge, but the inner edge is exponentially close to the horizon.





\section{Conclusion and discussion}
\label{condis}

The existence of a hairy AdS electron star solution as we have
constructed in this paper shows that relativistic critical theories
can support a mixed phase of bosons and fermions. Bosons don't always
win in these set-ups, and we have co-existence rather than
competition. The construction clearly shares a lot of
characteristics with 
the conventional counterexample of relativistic heavy
bosons with light fermions. In the bulk it is almost exactly what is
taking place, whereas in the boundary conformal field theory the role
of mass is played by the conformal dimension. 
One could have thought that 
the other simple counterexample, a very large number of fermion
flavors such that one can circumvent the Pauli principle, is also at play as we know that the fluid approximation in
the bulk essentially corresponds to a system with an infinite number
of quasiparticles distinguished by their AdS radial quantum number. It
is not clear that this is case. Based on our understanding of single
Fermi surface holography  \cite{Sachdev:2011ze,{Cubrovic:2010bf},{Allais:2012ye}, {Allais:2013lha}}  
most
of the above story will apply in that situation as well. 
Moreover including $1/N$ corrections in the bulk should render the
higher order Fermi surfaces unstable while keeping the macroscopic
characteristics of the star. 
Another criticism which one might raise is that it appears we are simply
describing a non-interacting system of bosons and fermions. A reason to think so is that 
for incommensurate charges where $q_b$ and $q_f$ are not integer multiples of each other one 
can think of the system as having a separate $U(1)_b$ and $U(1)_f$ symmetry, where we are keeping 
the off-diagonal chemical potential to zero.\footnote{We thank Steven Gubser for emphasizing this.}  
We argue
that this should also not apply on inspection. The direct argument is
that the field theory described is interacting, albeit rearranged in
a $1/N$ perturbative series, and similarly the AdS side clearly
experiences gravitational and electromagnetic
interactions. Preliminary results in a forthcoming companion paper where we study
the same system for $q_b=2q_f$ with a Yukawa interaction fully support
this view \cite{inprogress,{inprogress2}}.

\section*{Acknowledgments}
We thank Hong Liu for very useful correspondence. 
We would also like to thank  Steven Gubser and Richard Davison for helpful conversations. Y.L and Y.W. S would like to thank Rong-Gen Cai, Bin Chen, Yi Ling, Zheng-Yu Weng for discussions and they are very grateful to the hospitality of CHEP, Peking University and Shanghai Jiaotong University where this work was presented.
This research is supported in part by a Spinoza Award (J. Zaanen) from the Netherlands Organization for Scientific 
Research (NWO) and by the Dutch Foundation for Fundamental Research on Matter (FOM). 

%
%




\end{document}